\documentclass[12pt]{article}
\usepackage[font={it,small},justification=justified]{caption}
\usepackage{amsmath, mathpazo, amssymb, amsfonts, amsthm}
\usepackage[T1]{fontenc}
\usepackage{adjustbox}
\usepackage[utf8]{inputenx}
\usepackage{graphicx}
\usepackage{multirow}
\usepackage{dcolumn}
\usepackage{pdflscape}
\usepackage{natbib}
\usepackage{hyperref}
\newcolumntype{d}[1]{D{.}{.}{-1}}
\newcolumntype{s}[1]{D{,}{,}{-1}}
\usepackage[margin=.9in]{geometry}
\usepackage[normalem]{ulem}
\usepackage{enumitem}
\usepackage{xspace}
\usepackage[group-separator={,},group-minimum-digits=4]{siunitx}
\usepackage[hang,flushmargin]{footmisc}
\usepackage[toc, title]{appendix}
\usepackage{setspace}
\usepackage{bookmark}
\usepackage[svgnames]{xcolor}
\usepackage{hyperref}
\usepackage{booktabs}
\usepackage{comment}
\hypersetup{
colorlinks,breaklinks,naturalnames,
citecolor=DarkBlue,urlcolor=DarkBlue,linkcolor=DarkBlue
}
\usepackage{sectsty}
\sectionfont{\color{DarkRed}}
\subsectionfont{\color{DarkRed}}
\subsubsectionfont{\color{DarkRed}}
\usepackage{fancybox,fancyhdr}
\usepackage{mathtools}
\usepackage{bbm}
\usepackage{graphicx}
\usepackage{caption}
\usepackage{tikz}
\usepackage{subcaption}
\usepackage[thinc,text]{esdiff}
\usetikzlibrary{patterns}
\usetikzlibrary{shapes.geometric}
\usetikzlibrary{shapes.misc}
\usetikzlibrary{plotmarks}
\usepackage{mathabx}

\newlist{steps}{enumerate}{1}
\setlist[steps, 1]{label = Step \arabic*:}

\theoremstyle{plain}

\newtheorem*{theorem*}{Theorem}

\newtheorem{proposition}{Proposition}

\theoremstyle{definition}

\newtheorem{example}{Example}

\renewcommand{\epsilon}{\varepsilon}

\usepackage{array}
\newcommand{\PreserveBackslash}[1]{\let\temp=\\#1\let\\=\temp}
\newcolumntype{C}[1]{>{\PreserveBackslash\centering}p{#1}}
\newcolumntype{R}[1]{>{\PreserveBackslash\raggedleft}p{#1}}
\newcolumntype{L}[1]{>{\PreserveBackslash\raggedright}p{#1}}

\usepackage{clipboard}
\newclipboard{empirical-large-core}

\usepackage[capitalize,noabbrev,nameinlink]{cleveref}

\crefname{assumption}{assumption}{assumptions}
\usepackage{appendix}
\usepackage{etoolbox}
\AtBeginEnvironment{appendices}{\crefalias{section}{appendix}}
\newif\ifinappendix\inappendixfalse
\crefname{apptab}
  {\protect{\ifinappendix\else Appendix \fi}Table}
  {\protect{\ifinappendix\else Appendix \fi}Table}
\AtBeginEnvironment{appendices}{\inappendixtrue \crefalias{table}{apptab}}

\NewDocumentCommand{\expect}{ e{^} s o >{\SplitArgument{1}{|}}m }{%
  \operatorname{\mathbb{E}}
  \IfValueT{#1}{{\!}^{#1}}
  \IfBooleanTF{#2}{
    \expectarg*{\expectvar#4}%
  }{
    \IfNoValueTF{#3}{
      \expectarg{\expectvar#4}%
    }{
      \expectarg[#3]{\expectvar#4}%
    }%
  }%
}
\NewDocumentCommand{\expectvar}{mm}{%
  #1\IfValueT{#2}{\nonscript\;\delimsize\vert\nonscript\;#2}%
}
\DeclarePairedDelimiterX{\expectarg}[1]{[}{]}{#1}

\usepackage{titlesec}
\titlespacing*{\section}
{0pt}{2.5ex plus 1ex minus .2ex}{2ex plus .2ex}
\titlespacing*{\subsection}
{0pt}{2.25ex plus 1ex minus .2ex}{1.2ex plus .2ex}
\titlespacing*{\paragraph}
{0pt}{2.25ex plus 1ex minus .2ex}{1em}

\usepackage[most]{tcolorbox}

\makeatletter
\renewcommand\paragraph{\@startsection{paragraph}{4}{\z@}%
                                    {1.7ex \@plus1ex \@minus.2ex}%
                                    {-1em}%
                                    {\normalfont\normalsize\bfseries}}

\title{Human Misperception of Generative-AI Alignment:\\A Laboratory Experiment\thanks{We are grateful to Aislinn Bohren, Salvador Candelas, John Conlon, Krishna Dasaratha, Bnaya Dreyfuss, Tristan Gagnon-Bartsch, Alex Imas, Marc Kaufmann, Xiao Lin, Daniel Martin, Gali Noti, Yuval Salant, Chloe Tergiman, and Rui Wang for helpful comments.}}

\author{ \large Kevin He\thanks{University of Pennsylvania. Email: hesichao@gmail.com.} \ \ \ \  \ \ \ \  %
Ran Shorrer\thanks{The Pennsylvania State University. Email: rshorrer@gmail.com. Shorrer gratefully acknowledges support (for other projects) in the form of API credits from Anthropic, Google, and OpenAI.} \ \ \  \ \ \ \ \ %
Mengjia Xia\thanks{ University of Pennsylvania. Email: xiax@sas.upenn.edu.}}
\date{April 2, 2026}

\begin{document}

\maketitle

\begin{abstract}
\noindent We conduct an incentivized laboratory experiment to study people's perception of generative artificial intelligence (GenAI) alignment in the context of economic decision-making. Using a panel of economic problems spanning the domains of risk, time preference, social preference, and strategic interactions, we ask human subjects to make choices for themselves and to predict the choices made by GenAI on behalf of a human user. We find that people overestimate the degree of alignment between GenAI and human choices. In every problem, human subjects’ average prediction about GenAI's choice is substantially closer to the average human-subject choice than it is to the GenAI choice. At the individual level, different subjects' predictions about GenAI’s choice in a given problem are highly correlated with their own choices in the same problem. We explore the implications of people overestimating GenAI alignment using a simple theoretical model.
\end{abstract}

\section{Introduction}

Individuals and organizations are increasingly using generative artificial intelligence (GenAI) to help with their economic decisions.\footnote{For example,
GenAI is used in the context of financial trading \citep{cheng2025does,even2025effect},  hiring \citep{shrm2025talent,JAMA}, and pricing \citep{chafkin2025bloomberg}.
} This trend is accelerated by the rise of AI agents that can interact with the external environment and autonomously take actions on behalf of the user \citep{OpenAI_2025}, making it possible to even fully delegate economic decisions to GenAI.

Unlike classification and prediction tasks, where machine-learning methods and AI systems have been traditionally   deployed,  economic decisions often do not have an objectively ``correct'' answer that applies to everyone. Instead, these economic problems confront agents with trade-offs (e.g., higher payoff vs. earlier payoff, efficiency vs. equity, riskier rewards with a higher expectation vs. safer rewards with a lower expectation) and the optimal choices depend on the agent's preferences. To fully realize the potential gains from delegating  economic decisions to GenAI, users must hold correct beliefs about how this technology behaves when  instructed to act on their behalf. If users correctly anticipate GenAI's behavior, then judicious delegation of the appropriate decision problems to GenAI can save time and effort. But, as our theoretical analysis in \cref{sec:theory} shows, if users misperceive the degree of alignment between the GenAI choices and their own preferences, users may make suboptimal delegation decisions and even end up worse off than without access to GenAI.

This paper experimentally investigates the hypotheses that people overestimate the degree to which GenAI choices are aligned with human preferences in general (\emph{anthropomorphic projection}), and with their personal  preferences in particular  (\emph{self-projection}).\footnote{The hypotheses and our main analyses were pre-registered. The  pre-registration can be found on
the registry website at \href{https://aspredicted.org/yd32-r96n.pdf}{https://aspredicted.org/yd32-r96n.pdf}.} To this end, we conduct an incentivized laboratory experiment where we focus on understanding people's beliefs about GenAI's choices.\footnote{
A related growing literature focuses on studying the responses produced by large language models (LLMs) instead of people's beliefs about these models. Some of the work in this area considers the possibility of using LLMs to make predictions about humans \citep[e.g.,][]{gao2026llm} or simulate human subjects \citep[e.g.,][]{horton_large_2023,manning_automated_2024,GABE}, while others study LLMs as economic agents in order to understand how they behave in markets \citep[e.g.,][]{fish2024algorithmic,EconEvals,cook2026llms}.} The experiment consists of two parts. In the first part, subjects are asked to make choices in an array of incentivized decision environments spanning the domains of risk, time preferences, social preferences, and strategic interactions.\footnote{\citet{Yariv} study most of these decision environments and compare behavior across different human subject pools. Our subjects' choices are in line with their findings.} In the second part, subjects are asked to predict the choices an AI chatbot would make when instructed to choose on behalf of a human user in the same decision environments. Subjects receive a bonus  if their prediction is sufficiently close to the average choice made by the large language model (LLM) GPT-4o.

We find evidence of both anthropomorphic projection and self-projection.  First, on average, human subjects' predictions about GenAI's choices in every decision environment are much closer to the average human-subject choice than to the average GenAI choice. Second, at the individual level, human subjects' predictions about GenAI's choices in a given environment are highly correlated with their own choices in the same environment. Additionally, consistent with subjects self-projecting preference parameters and not just specific choices onto the GenAI model, we find that a subject’s expectation of how GenAI would choose for a human user in a given problem can be predicted from the subject’s choices in related problems.

Our analysis is informed by a simple theoretical model of user perception of GenAI that accommodates anthropomorphic projection and self-projection. Using this model, we show that both types of projection coexist. Namely, our data  reject the hypotheses that people do not exhibit anthropomorphic projection or do not exhibit self-projection.  We also use the model to explore the implications of anthropomorphic projection and self-projection in a stylized delegation setting. We find that these misperceptions can lead to over-delegation to GenAI. More subtly,  objectively improving AI alignment can harm agents who exhibit anthropomorphic projection (because they mistakenly adjust their delegation decisions in a detrimental fashion). Similarly, among agents who exhibit self-projection, welfare may be higher for those who have more unusual preferences (since they are less likely to mistakenly delegate). These results contribute  to the literature on delegation to AI \citep[e.g.,][]{fudenberg2026friend,liang2026clones} and to the broader literature on AI alignment \citep[e.g.,][]{gabriel2020artificial,hosseini2025} by analyzing the implications of \emph{misperceptions} of alignment when people selectively delegate to GenAI.

Anthropomorphic projection and self-projection may result from several causes.  First, humans may believe that GenAI models are designed to behave like humans, and a large literature documents that people project their current tastes and knowledge onto other people when they forecast others’ behavior and studies some implications of this bias \citep{danz2018biases,gagnon2021projection,bushong2024failures,gagnon2024quality}.\footnote{\citet{rubinstein2016isn} present evidence of ``self-similarity'' in symmetric two-player games: players that choose an action $X$ attribute higher probability for other choosing $X$.} Furthermore, one may expect excessive projection even when people interact with personalized GenAI models, as the literature  shows that people also project their current tastes (which may be influenced by contextual information that the GenAI cannot observe or interpret) onto their future or past selves \citep[e.g.,][]{loewenstein2003projection,conlin2007projection,kaufmann2022projection}. Second, predicting the choices of GenAI in a specific environment is difficult, especially for individuals with less experience with GenAI products, and this may lead people to rely on a simple cognitive default \citep{woodford2020modeling}.

To assess the possibility that experience mitigates self-projection,  we collect information on subjects' exposure to GenAI and conduct heterogeneity analysis. We find no evidence that the extent of self-projection varies substantially by past experience with GenAI. Additionally, we find no evidence that experienced subjects make more accurate predictions about GenAI's choices. To compare self-projection onto GenAI with projection bias onto other humans, we replicate our main experiment, asking subject to predict other human subjects' choice. We find that the magnitude of self-projection onto GenAI is comparable in magnitude to projection onto other humans.\footnote{By contrast, subjects' predictions of other humans' choices are substantially more accurate than their predictions of GenAI choices.}

We view our primary innovation as documenting self-projection, which may be more robust to future developments in GenAI. Specifically, we find that subjects' individual predictions of GenAI choice tend towards their own preferred choices in the same problem. This pattern is independent of the actual GenAI choice. We think that this bias is likely to be present whenever people with heterogeneous preferences make predictions about the same ground-truth GenAI response to the same prompt, independent of the quality or alignment of that GenAI response.

On anthropomorphic projection, we think of our contribution as pointing out the theoretical implications of this bias and showing that such misperception exists to a significant extent for the state-of-the-art GenAI models (at the time of our study) and for a selection of canonical economic problems. New models in the future may well become more aligned with people’s average perception of how they behave. We hope that our paper can lead to future work that will measure the extent of anthropomorphic projection in newer models and for other economic decision problems.

\subsection{Related Literature}

Our paper is closely related to studies that consider humans' belief formation about AI ability and their decision to delegate to AI.
\citet{vafa_large_2024} and \citet{dreyfuss2024human}
provide evidence that humans make anthropomorphic generalizations about LLM behavior in questions that involve factual answers. \citet{vafa_large_2024} show that when asked to guess how an agent will perform in one task based on the agent's performance in another task, human subjects do well when the agent is human, but they perform poorly when the agent is an LLM. \citet{dreyfuss2024human} show that human subjects project onto the LLM a notion of human difficulty and capability, even though it does not apply to the LLM.\footnote{In the field of human-computer interaction, studies in the celebrated Computers Are Social Actors (CASA) literature have shown that humans apply social response patterns to technology despite recognizing that technological systems are not human \citep{reeves1996media,nass2000machines}.} \citet{dell2023navigating}  coin the term ``jagged technological frontier'' to describe how GPT-4 performs well in some tasks but poorly in other seemingly similar tasks. They show that giving professional management consultants access to GPT-4 can be detrimental to their performance when the task is on the wrong side of the technological frontier.\footnote{There are ample evidences that LLMs can augment performance in a variety of tasks \citep[e.g.,][]{brynjolfsson2025generative}. }
\citet{Noti} design an AI system that provides advice only when it is likely to be beneficial for the user and show that it can improve human decision-making relative to a design that always provides advice. In a similar spirit, \citet{ruru} propose coarsening the AI signal space to account for human users' cognitive biases. We contribute to this literature by considering economic decision environments where agents' optimal choices vary based on their preference parameters. This setting lets us document a novel, distinct phenomenon: self-projection.

More broadly, our findings contribute to several strands of academic research. First, they contribute to the vast literature on mental models in decision-making.\footnote{E.g., \citet{mullainathan2008coarse}, \citet{bordalo2012salience}, \citet{bordalo2016stereotypes}, \citet{enke2019correlation}, \citet{imas2022impact}, and \citet{esponda2024mental}.} Second, they contribute to the growing literature on the interaction of algorithms with society.\footnote{E.g.,  \citet{calvano2020protecting}, \citet{rambachan2020economic}, \citet{aquilina2022quantifying}, \citet{liang2022algorithmic}, and \citet{liang2026creative}. } Finally, they contribute to the Human+AI literature.\footnote{E.g., \citet{kleinberg2018human}, \citet{immorlica2024generative}, \citet{mullainathan2025economics}, \citet{noti2025ai}, and \citet{imas2025agentic}. }

\section{Theoretical Framework}
\label{sec:theory}

In this section, we present a stylized theoretical model of anthropomorphic projection and self-projection.  We then apply this model to a delegation setting and derive  comparative statics of GenAI users' welfare under these misperceptions. Later, in Section \ref{sec:main_results}, we will use this model to interpret our experimental results.

\subsection{A Model of Anthropomorphic Projection and Self-Projection}
\label{subsec:projection_model}
There is a population of agents, each with a type $\theta$ drawn from the distribution $F_\theta$.
Each agent must take an action $a \in \mathbb{R}$ in a decision problem $\omega \in \mathbb{R}$. The agent's   optimal choice given the decision problem and  their type is $a(\omega,\theta)$. The average ideal action within the agent population for decision
problem $\omega$ is $\mathbb{E}_{\theta}[a(\omega,\theta)]$.

When GenAI is asked to choose an action on behalf of an agent (of any type) in decision problem $\omega$, GenAI
takes the action $y(\omega)$ that only depends on $\omega$. We define $b(\omega) := y(\omega) - \mathbb{E}_{\theta}[a(\omega,\theta)]$, so we can decompose $y(\omega) = \mathbb{E}_{\theta}[a(\omega,\theta)]+b(\omega)$.
We interpret the term $b(\omega)$
as the bias of the GenAI relative to the population of agents for decision problem
$\omega$. We are agnostic about the source of such bias (for instance, biased training sample or issues with the model-training procedure as in \citealp[]{autor2025misaligned}) and allow the amount of bias to depend on the decision problem in an arbitrary way.

In practice, delegation to GenAI may lead to a partially personalized action that depends on the delegator's type. This may be because people choose to use one of several available GenAI models depending on their personal type realizations, or because the GenAI model has access to the agent's personal information and tailors its choice based on this information. We can view $\theta$ as the remaining idiosyncratic preference or contextual information that is orthogonal to the GenAI personalization \citep{iakovlev2025value}.

Agents potentially misperceive GenAI's action. An agent of type $\theta$ thinks that GenAI will take the action  $\rho\cdot a(\omega,\theta)+(1-\rho)\mathbb{E}_{\theta}[a(\omega,\theta)]+r\cdot b(\omega)$ in decision problem $\omega$, for $0\le\rho,r\le1.$ Agents act optimally given their misperceptions.

In our interpretation, $\omega$ contains the specific details of a decision
to be made, such as the rate of return on a risky investment or the
social benefit of a generous act. The agent's optimal action depends on both the decision problem $\omega$ and their type $\theta$, which
refers to a  personal trait such as risk attitude or social-preference
parameter. We assume that the agent knows their type, but whether the agent knows the decision problem depends on the application.

The model accommodates both  anthropomorphic projection and  self-projection. The parameter
$r$ relates to anthropomorphic projection, where agents on average predict the GenAI action in problem $\omega$
to be $\mathbb{E}_{\theta}[a(\omega,\theta)]+r\cdot b(\omega)$. Thus, when GenAI is biased (i.e., $b(\omega)\ne 0$), anthropomorphic projection becomes more severe as $r$ decreases, with people's  predictions of GenAI's action becoming more centered around the average ideal human action and further away from  the actual GenAI action in each decision problem. The parameter $\rho$ refers to the extent of self-projection, where agents partially project their own
individual type onto the GenAI. Notably, $\rho$ has no effect on humans' average prediction of GenAI action.

As we will show below, our estimation procedures will identify the $r$ and $\rho$ parameters using experimental data. To interpret these parameters, we first consider a few special cases.

\paragraph{Correctly specified beliefs.} In this case, the agent  fully anticipates  the GenAI bias ($r=1$), and does not project their own preferences onto GenAI ($\rho=0$).

\paragraph{No anthropomorphic projection, some self-projection.} In this case, the agent fully anticipates the GenAI bias ($r=1$), but projects their own preferences onto GenAI ($\rho>0$).

\paragraph{No self-projection, some anthropomorphic projection.} In this case, the agent does not project their own preferences onto GenAI ($\rho=0$), but does not fully anticipate the GenAI bias ($r<1$).

\paragraph{Weighted average of self and GenAI.} In this case,  the agent simply perceives the GenAI's
action as a $\phi$-convex combination between their optimal choice and the choice that the GenAI actually makes. So, type $\theta$  perceives GenAI's action to be
\begin{equation}\label{eq: convex preferences}
    \phi\cdot a(\omega,\theta)+(1-\phi)\cdot(\mathbb{E}_{\theta}[a(\omega,\theta)]+b(\omega))=\phi\cdot a(\omega,\theta)+(1-\phi)\mathbb{E}_{\theta}[a(\omega,\theta)]+(1-\phi)\cdot b(\omega).
\end{equation}
\cref{eq: convex preferences} reveals that the weighted-average model corresponds to the special case of our model  where
$\rho+r=1.$

\subsection{Application: Delegation under Misperceived Alignment}

Our experiment will provide evidence of anthropomorphic projection and self-projection. Here, we study the implications of these misperceptions in a delegation setting, motivated by the recent developments in \emph{agentic AI}:  autonomous AI agents that can take actions on behalf of their users. We study an environment where the decision problem $\omega$ is not transparent to the agents, so they must pay an attention cost in order to learn $\omega$. In this scenario, agents may find it beneficial to delegate decision to GenAI to save on attention cost, but we show how such delegation decisions can be distorted by agents' misperceptions about GenAI alignment.

Nature draws a decision problem $\omega\sim\mathcal{N}(0,\sigma_{\omega}^{2})$, which is not observed by the agent.
The agent observes their type $\theta\sim\mathcal{N}(0,\sigma_{\theta}^{2})$
and an attention cost $c>0$ , where $c$ is drawn from a strictly
positive density on $\mathbb{R}_{+}$ (and is independent of $\theta$ and  $\omega$). An action  $a \in \mathbb{R}$ must be taken and the agent with type $\theta$ gets decision utility $-(a-\omega-\theta)^{2}$ from action $a$ in decision problem $\omega$. Mapping back to the general setup, the optimal action as a function of the decision problem and type is $a(\omega, \theta)=\omega+\theta$.

The agent first chooses whether to costlessly
delegate their action to the GenAI. When the decision problem is $\omega$
and the agent delegates, the GenAI will take the action  $\omega+b(\omega)$
on behalf of the agent (regardless of the agent's type $\theta$).
If the agent does not delegate, then they must choose an action themselves. Before doing so, they have the chance to  pay the attention cost $c$ to acquire a signal $s\sim\mathcal{N}(\omega,\sigma_{s}^{2})$ about the decision problem $\omega$.\footnote{This information-acquisition process takes time and the agent loses the opportunity to delegate to GenAI if they choose to acquire the signal $s$.} If the agent does not delegate and does not pay the attention cost, then they must choose an action knowing only their type $\theta$.

The agent acts optimally given their misspecified beliefs about  the GenAI's action. As in \cref{subsec:projection_model},  a type $\theta$ agent believes that the GenAI
will take the action $\omega+rb(\omega)+\rho\theta$ in decision problem
$\omega,$ for some $0 \le r,\rho \le 1.$ The agent maximizes expected total utility (i.e., decision utility minus any attention cost) given these beliefs.

The interpretation is that the agent knows the distribution of decision problems in the world, but must pay a cost $c>0$ to investigate the details of the particular problem that they are currently facing. For example, someone who is asked to donate to a  charitable organization may know the distribution of effectiveness levels among all charities, but they must research the financial statements and track record of this specific organization to understand where it lies in the distribution. The magnitude of the signal variance $\sigma_s^2 \ge 0$ captures the noise in the agent's information when they choose to learn about the decision problem. In problems with a large $\sigma_s^2$, agents find it so difficult to understand the problem's details that GenAI chooses much more accurate actions than human agents relying on the noisy signal, despite the GenAI's bias and  lack of personalization.

\subsubsection{When Do Agents Delegate?}

When the agent takes an action, they maximize their expected decision utility by choosing an action equal to their type plus the mean of their current belief about $\omega$. A type $\theta$ agent who does not delegate and does not acquire a signal maximizes decision utility by choosing the ex-ante optimal action, $a=\theta$.
By the Gaussian updating formula, an agent who sees a signal realization $s$ has an updated posterior belief about $\omega$ whose mean is $\frac{\sigma_{\omega}^{2}}{\sigma_{\omega}^{2}+\sigma_{s}^{2}}\cdot s$. The agent's posterior variance about $\omega$ is the same for every signal realization. Given the signal $s$, the utility maximizing action for an agent of type $\theta$ is $\frac{\sigma_{\omega}^{2}}{\sigma_{\omega}^{2}+\sigma_{s}^{2}}\cdot s + \theta$, and the agent's expected decision utility does not vary with the realization of $s$.

The optimal strategy of an agent with type $\theta$ must be one of
the following: (1) choose the ex-ante optimal action $a=\theta$ without
delegating to GenAI and without learning; (2) acquire a signal $s$
and choose the action $\frac{\sigma_{\omega}^{2}}{\sigma_{\omega}^{2}+\sigma_{s}^{2}}\cdot s+\theta$;
(3) delegate to GenAI. The expected utility associated with the first
two strategies are $-\sigma_{\omega}^{2}$ and $-(c+\frac{\sigma_{s}^{2}\sigma_{\omega}^{2}}{\sigma_{s}^{2}+\sigma_{\omega}^{2}})$
and they do not vary with the agent's perception of GenAI.

If the type $\theta$ agent perceives the GenAI action in problem
$\omega$ as $\omega+rb(\omega)+\rho\theta$, then their \emph{subjective}
expected utility from the third strategy is
\begin{align*}
-\mathbb{E}[((\omega+rb(\omega)+\rho\theta)-(\omega+\theta))^{2}] & =-\mathbb{E}[(rb(\omega)+(\rho-1)\theta)^{2}].\\
 & =-\left(r^{2}\mathbb{E}[b(\omega)^{2}]+(1-\rho)^{2}\theta^{2} -2(1-\rho)r\theta\mathbb{E}[b(\omega)] \right)
\end{align*}
Here, $r^{2}\mathbb{E}[b(\omega)^{2}]$ is the agent's expected loss
due to GenAI's bias relative to the average optimal human action,
$(1-\rho)^{2}\theta^{2}$ is the agent's expected loss due to the
GenAI not tailoring its action to the agent's type realization $\theta.$
The third term $-2(1-\rho)r\theta\mathbb{E}[b(\omega)]$
can mitigate expected losses if the agent's type realization $\theta$
matches the average bias of the GenAI. For instance, if the GenAI
tends to take higher actions than the optimal action for the average
human, then  above-average types are better off relative to lower-than-average types.

Now suppose $\mathbb{E}[b(\omega)]=0$. So, even though the GenAI
may be biased on every particular problem, the bias is not systematic across
problems. Then the foregoing discussion establishes that:

\begin{proposition}
    When $\mathbb{E}[b(\omega)]=0$, the agent strictly prefers to delegate
to GenAI if and only if
\begin{equation}\label{eq: conditions}
    r^{2}\mathbb{E}[b(\omega)^{2}]+(1-\rho)^{2}\theta^{2}<\min\left\{\sigma_{\omega}^{2},c+\frac{\sigma_{s}^{2}\sigma_{\omega}^{2}}{\sigma_{s}^{2}+\sigma_{\omega}^{2}}\right\}.
\end{equation}
\end{proposition}
 The left hand side of \cref{eq: conditions} decreases when $r$ gets smaller and when $\rho$
grows larger, so increasing anthropomorphic projection or self-projection (weakly) increase the probability of delegation. The left hand side is increasing in $|\theta|$, so agents
with more atypical preferences are less likely to delegate to GenAI.
The effect of the agent's signal accuracy appears through $\sigma_{s}^{2}.$
Fixing the other parameters, for large enough $\sigma_{s}^{2}$ the
strategy of acquiring a signal and choosing the action $\frac{\sigma_{\omega}^{2}}{\sigma_{\omega}^{2}+\sigma_{s}^{2}}\cdot s+\theta$
is suboptimal and so the agent will choose to either take
the ex-ante optimal action or to delegate to GenAI.

In the following two sections we separately consider the effects of anthropomorphic projection and self-projection on agent's delegation behavior and welfare. For the sake of clarity, we assume that if the agent pays the attention cost $c$, then they observe a perfectly informative signal (i.e., $\sigma^2_s =0$).

\subsubsection{Implications of Anthropomorphic Projection}
Suppose $\sigma_{\theta}^{2}=0$, so there is no individual-level
variance in optimal actions. We show that anthropomorphic projection  causes over
delegation to the GenAI.

\begin{proposition}\label{prop:group_level}
There is a threshold $\bar{r}\in[0,1]$ so  that when $r>\bar{r}$,  the agent never delegates to GenAI and behaves in the same way as a correctly specified agent. When
$r\le\bar{r}$, the agent delegates to GenAI when $c>r^{2}\mathbb{E}[b(\omega)^{2}]$
and pays the attention cost when $c<r^{2}\mathbb{E}[b(\omega)^{2}]$,
and  the probability of over-delegation is  strictly decreasing in $r$
over the range $[0,\bar{r}].$ The threshold $\bar{r}$ is strictly
interior when $\mathbb{E}[b(\omega)^{2}]>\sigma_{\omega}^{2}$ and
it is equal to 1 when $\mathbb{E}[b(\omega)^{2}]<\sigma_{\omega}^{2}$.
\end{proposition}

In the case where the GenAI's bias is relatively large ($\mathbb{E}[b(\omega)^{2}]>\sigma_{\omega}^{2}$),
a correctly specified agent never delegates to GenAI. Instead, the agent
either pays the attention cost to learn $\omega$ when $c$ is low
enough, or chooses the ex-ante optimal default action 0 when $c$
is too high. With sufficiently severe anthropomorphic projection,
the biased agent over delegates. For high $c,$ the biased agent delegates
to GenAI while the correctly specified agent chooses the default action. For medium
$c,$ the biased agent delegates to GenAI while the correctly specified agent pays
the attention cost.

Even in the case where the GenAI's bias is relatively small ($\mathbb{E}[b(\omega)^{2}]<\sigma_{\omega}^{2}$)
so that a correctly specified agent sometimes delegates to GenAI, the biased agent
still uses a wrong threshold in cost realization to decide
between paying attention or delegating to GenAI.
For some intermediate values of $c$, a correctly specified agent pays attention
but the biased agent delegates.

A corollary  of \cref{prop:group_level} is that an agent who suffers from anthropomorphic projection can be made strictly worse off when the GenAI becomes objectively more aligned on every problem. Of course, this cannot happen to a correctly specified agent, and it also cannot happen under any fixed (even if irrational) delegation strategy that maps attention cost realizations to delegation decisions. As the following example illustrates, this phenomenon happens because the biased agent increases their GenAI delegation by too much in response to the GenAI's improved alignment, and this behavioral adjustment in delegation is what drives down their welfare.

\begin{example}
Fix any $0<r<1$ and consider $b_{L}=(\sigma_{\omega}/r)-\varepsilon$
and $b_{H}=(\sigma_{\omega}/r)+\varepsilon$ for sufficiently small $\varepsilon>0$
so that we still have $b_{L}>\sigma_{\omega}$. Consider a GenAI model with
$b(\omega)=b_{H}$ for every $\omega$ and another GenAI model with $b(\omega)=b_{L}$
for every $\omega$. For a correctly specified agent, because both $b_{L}^{2}$
and $b_{H}^{2}$ are larger than $\sigma_{\omega}^{2}$, Proposition
\ref{prop:group_level} implies the correctly specified agent never delegates to either GenAI model and has
the same welfare when they have access to either. By contrast, the biased
agent with parameter $r$ does not delegate for $b(\omega)=b_{H}$
(and gets the same welfare as the correctly specified agent) but delegates with
positive probability for $b(\omega)=b_{L}$ (and gets strictly lower
welfare compared to the correctly specified agent since they are always strictly better off choosing
$a=0$ instead of delegating). So, the biased agent has strictly lower
welfare when they have access to a GenAI model with the lower bias $b(\omega)=b_{L}$
than a GenAI model with the higher bias $b(\omega)=b_{H}$.
\end{example}

\subsubsection{Implications of Self-Projection}

Suppose $b(\omega)=0$ for every $\omega$, so the GenAI takes the
optimal action for the average agent in every decision problem (we maintain the assumption that $\sigma^2_s =0$). If
an agent  exhibits self-projection
bias with $\rho=1$, then they believe that the GenAI will take their
optimal action in every decision problem. So, they will make the mistake
of always delegating their decisions. The next proposition generalizes this special case: under any amount of self-projection, agents over-delegate to GenAI (strictly if their type $\theta$ is not too extreme) because they over-estimate the degree to which the
AI's decision matches their idiosyncratic preferences.

\begin{proposition}
\label{prop:self_basic}
Suppose $\rho\in[0,1).$ For $|\theta|>\sigma_{\omega}/(1-\rho)$,
the agent  never delegates to GenAI for any
realization of $c$ and behaves as-if they are correctly specified.  For $\sigma_{\omega}<|\theta|<\sigma_{\omega}/(1-\rho)$,
the correctly specified agent never delegates to GenAI but the biased agent delegates
to GenAI with positive probability. For $|\theta|<\sigma_{\omega}$, both the correctly specified agent and the
biased agent delegate to GenAI with positive probability, but the
biased agent does so for more realizations of the attention cost $c$.
\end{proposition}

The idea behind this result is that an agent with type $\theta$ partially
projects their type onto the GenAI's behavior, thus misperceiving the
expected decision utility from delegation to be $-(1-\rho)^{2}\theta^{2}$
instead of the objectively correct $-\theta^{2}$. This causes the
agent to over-delegate compared to the correctly specified benchmark.

For correctly specified agents in a world with GenAI, welfare is monotonically decreasing
in the distance of an agent's type to the group average. The intuition
is that the GenAI is more aligned with the average agent but less aligned
with agents with more unusual preferences, so the option of delegation
is less beneficial for the latter. But this result crucially depends
on agents holding correct beliefs about the GenAI behavior and need not hold when
agents suffer from self-projection. Indeed, for $\rho\in(0,1),$ we
show that welfare jumps up discontinuously at the type $\theta=\sigma_{\omega}/(1-\rho)$.
The idea is that a biased agent who is subjectively almost indifferent between
delegating to GenAI and taking the default action equal to their type
is actually substantially better off taking the default action, since
the subjective indifference is driven by an overestimation of the
alignment between the GenAI's action and the agent's type.
\begin{proposition}
\label{prop:monotonic}
For correctly specified agents who can delegate to GenAI, welfare is monotonically
decreasing in $|\theta|.$ By contrast, for $\rho\in(0,1),$ the welfare
of agents who suffer from self-projection is not monotonic around $\theta=\sigma_{\omega}/(1-\rho)$.
\end{proposition}

\section{Experimental Design and Deployment}
\label{sec:design}

\subsection{Overview}

We advertised the experiment as a study that requires people to make  incentivized choices and predictions. The experiment began with a brief informed consent. Subjects who consented were told that the experiment consists of two parts, that they will earn ``tokens'' based on their answers, and that these tokens will be converted into a bonus payment at a rate of 1,000 tokens per US dollar at the end of the experiment (in addition to a base payment).

In the first part of the experiment (\emph{choice tasks}), subjects are asked
to make choices in nine problems spanning four domains: risk, time
preference, social preference, and strategic interactions (see \cref{subsec:The-Decision-Problems} for details on the problems). Problems appear in a random order: for each subject,
we draw uniformly at random an order of the four domains, and within each domain we  also randomize the
order of the problems. Subjects earn tokens based
on their choices in every problem. Subjects receive no feedback during the experiment (specifically, they only learn how much they earned  after the end of the experiment).

In the second part of the experiment (\emph{prediction tasks}),
subjects are told that an AI chatbot was asked to make  choices on the behalf of a human user in the same problems (and in an additional problem that the subjects have not seen before). They are also shown the exact instructions that were given to the chatbot before each choice:

\begin{quote}
``You are a powerful decision-making agent and a helpful assistant that strictly follows the user’s instructions. The user is busy and requires you to provide an answer in exactly the requested format. The user may be given tokens depending on the answer you provide; each token is worth 0.001 US dollars. Here is the question that the user is facing:''
\end{quote}

Subjects are told that the AI chatbot was asked about each problem thousands of times, and they are asked to predict the average AI response. The problems in the prediction tasks appear in a random order according to the same procedure we used in Part 1 (but using an independent random draw).  Subjects earn 100 tokens for each
prediction task where their prediction is sufficiently accurate (no more than 10\% off from
the average AI choice).

\subsection{\label{subsec:The-Decision-Problems}The Decision Problems}

We assembled a panel of ten economic decision problems across the four domains: risk, time preference, social preference, and strategic interactions. Eight of the problems came from \citet{Yariv}, who use these (and other) tasks to compare behavior across different experimental subject pools. We added an additional problem of strategic interaction (the beauty contest, Decision Problem 9) and an additional problem of social preference (Decision Problem 10) that uses slightly different numbers than those used in \citet{Yariv}. Each problem requires either a numerical answer or a binary answer.

\paragraph{Decision Problem 1 (``risk100'').} The subject chooses how many tokens to wager out of an endowment of 100.  With 35\% probability, the subject receives three times the wagered tokens. With 65\% probability, the wagered tokens are lost.\footnote{As in \citet{Yariv}, decision problems in the domain of risk follow \citet{gneezy1997experiment}. }

\paragraph{Decision Problem 2 (``risk200'').}
The subject chooses how many tokens to wager out of an endowment of 200.  With 50\% probability, the subject receives 2.5 times the wagered tokens. Otherwise, the wagered tokens are lost.

\paragraph{Decision Problem 3 (``discounting'').} Subjects will receive either 150 tokens in 30 days or a larger number of tokens in 60 days. They are asked to report the minimal number of tokens that will make them choose the 60-days option. The number of tokens associated with the 60-days option is then randomly drawn from the interval between
150 and 400, and the subject receives the option that matches their reported threshold.\footnote{This problem is adapted from \citet{Yariv}, who use a similar but
hypothetical comparison between money in 30 days versus 60 days. We
chose the range of 150 tokens to 400 tokens for the 60-days option
based on their finding that the vast majority of
subjects give answers in this range.}

\paragraph{Decision Problem 4 (``dictator100'').}
The subject chooses how many tokens, out of an endowment of 100, to give away to another randomly selected subject.

\paragraph{Decision Problem 5 (``dictator300'').}
The subject chooses how many tokens, out of an endowment of 300, to give away to another randomly selected subject.

\paragraph{Decision Problem 6 (``dictator100x2'').}
The subject chooses how many tokens, out of an endowment of 100, to give away to another randomly selected subject. For each token given, the other subject receives two tokens.

\paragraph{Decision Problem 7 (``dictator100x0.5'').}
The subject chooses how many tokens, out of an endowment of 100, to give away to another randomly selected subject. For each token given, the other subject receives half a token.

\paragraph{Decision Problem 8 (``prisoner'').} Subjects play a one-shot
prisoner's dilemma  with another randomly selected subject from the same session.
If both players cooperate, then each gets 80 tokens. If one cooperates
and one defects, then the cooperator gets 60 tokens and the defector
gets 90 tokens. If both defect, then each gets 70 tokens. The two
actions in the game are given abstract names to avoid any connotations
of the words ``cooperate'' and ``defect.''

\paragraph{Decision Problem 9 (``beauty'').} Subjects play ``guess two-thirds
the average,'' an instance of a beauty-contest game. Subjects enter
whole numbers between 0 and 100, and the subject whose number is closest
to two-thirds of the average of the numbers entered by all subjects
in the session wins 5,000 tokens.

\paragraph{Decision Problem 10 (``dictator200'').}
The subject chooses how many tokens, out of an endowment of 200, to give away to another randomly selected subject. This problem was not presented to the human subjects as a choice task, but they were
asked to make a prediction about GenAI's choice in this problem during the prediction tasks (Part 2 of the experiment).

For easy reference,  \cref{tab:problems_summary} summarizes the descriptions of the decision problems.

\begin{table}
\begin{centering}
\caption{\label{tab:problems_summary}Summary of Decision Problems}
\begin{tabular}{>{\raggedright\arraybackslash}m{2.5cm}m{3cm}m{11cm}}
\toprule \textbf{Domain} &
\textbf{Task} & \textbf{Description}\tabularnewline
\midrule
\multirow{2}{=}[-15pt]{{Risk\\Preference}} &
risk100 & Wager some of 100 tokens: 35\% chance to receive 3 times the wagered
tokens, 65\% chance to lose them.\tabularnewline\tabularnewline
&
risk200 & Wager some of 200 tokens: 50\% chance to receive 2.5 times the wagered
tokens, 50\% chance to lose them.\tabularnewline

 \midrule

\multirow{1}{=}[8pt]{{Time\\Preference}} &
discounting & A delayed payment of 150 tokens in 30 days would be equivalent to a delayed payment of how many tokens in 60 days for you? \tabularnewline
\midrule
\multirow{5}{=}[-35pt]{{Social\\Preference}} &
dictator100 & Give tokens to a random subject from an endowment of 100.\tabularnewline\tabularnewline
&
dictator200 & Give tokens to a random subject from an endowment of  200.\tabularnewline\tabularnewline
&
dictator300 & Give tokens to a random subject from an endowment of  300.\tabularnewline\tabularnewline
&
dictator100x2 & Give tokens to a random subject from an endowment of  100. Recipient gets 2 tokens for each token given away. \tabularnewline\tabularnewline
&
dictator100x0.5 & Give tokens to a random subject from an endowment of  100. Recipient gets half a token for each token given away.\tabularnewline
 \midrule
 \multirow{2}{=}[-12pt]{{Strategic\\Interactions}} &
prisoner & Choose cooperate or defect in a prisoner's dilemma game.\tabularnewline\tabularnewline
&
beauty & Choose a number between 0 and 100 in a  beauty-contest game (guess two-thirds of the average guess).\tabularnewline
\bottomrule
\end{tabular}
\par\end{centering}

\end{table}

\subsection{Querying GPT-4o}
The GenAI choices used to evaluate the correctness of subjects' predictions both for payment and for the main analysis were obtained from GPT-4o. We designed our prompts so that the GenAI model outputs a choice as the first token without offering detailed reasoning steps (see \cref{sec:prompts} for details). OpenAI provides the log probabilities for up to the 20 most likely tokens at each position. Accordingly, we recorded the log probabilities of the top 20 tokens at the first position and calculated a weighted average with weights proportional to their probabilities.\footnote{The probabilities of the top 20 most likely tokens added up to $0.996$ on average.} The only exception is the prisoner's dilemma, which requires the GenAI models to make a binary choice between the two strategies ``A'' (cooperate) and ``B'' (defect). In this case, we specifically recorded the probability of token ``A.'' Since log probabilities are not fully deterministic, we repeated this process 100 times and took the average as the final choice.
\subsection{Deployment}
We implemented the experiment in oTree \citep{otree} and conducted it online using the Prolific platform in January  2025. We recruited 300 subjects who met the following three criteria: (1) live in the United States; (2) have previously
completed at least ten studies on Prolific; (3) have an approval rate
of at least 95\% on Prolific. Subjects were recruited in three sessions, with 100 subjects per session.
Subjects had up to 67 minutes to complete the study. On average, they took 12.97 minutes (s.d. 10.05 minutes) and earned \$4.15 (s.d. \$0.59), including a show-up fee of \$2.70.\footnote{A small part of this payment was delayed by 30 days or 60 days due to Decision Problem 3, which elicits time preferences. See \cref{subsec:The-Decision-Problems} for details.} Thus, the average earning rate was \$19.20 per hour.

\subsection{Auxiliary Measures and Questions}

At the end of the study, we asked subjects several questions about their degree of exposure, usage intensity, and attitudes towards GenAI (see \cref{fig:survey_results}).
We also have access to demographic data on the subjects from their Prolific account registration.

In addition, throughout the experiment, we tracked  the amount of time that subjects spent on each task (choices and predictions). To mitigate the risk that subjects use LLMs in prediction tasks,  we also kept track
of subjects who copied text from the webpage during the tasks. Specifically,
subjects who pressed the keyboard combination Ctrl+C on Windows, Command+C on Mac, or used the copy function in their web browser during a task are flagged in our data. We found that 11\% of the subjects copied text at least once.

Finally, for robustness, and since subjects were not informed of the specific GenAI model used in the prediction tasks, we also queried three additional commercial models (GPT-4o-mini,\footnote{On average, the probabilities of the top 20 most common  tokens from the GPT-4o-mini model added up to $0.995$.} Gemini-1.5-Pro, and Gemini-1.5-Flash). The prompts provided to each model were identical, although the methods for eliciting choices varied. Specifically, since Gemini models do not provide the distribution of the next token, we queried Gemini models 1,000 times for each task and computed the average result. These measurements were used for supplemental analyses, but not for determining subjects'
compensation.

\subsection{Pre-Registration}

We pre-registered our experimental protocol and primary analyses prior
to the start of the experiment. Our pre-registration specifies GPT-4o
as the model to be used to test the accuracy of subjects' predictions, the target sample size (300), a measure of the relative accuracy of aggregate subject predictions
about the GenAI choices (see \cref{subsec:PRA}), a regression specification to estimate individual-level
self-projection (see \cref{subsec:self-projection}), and a similar regression specification with the subject's
prediction for a particular problem as the dependent variable and
the subject's choice in a related problem as the regressor. The
pre-registration also discussed our secondary analyses relating to subjects' experience with  and attitudes towards
GenAI tools, but we did not specify any particular hypotheses. The  pre-registration can be found on
the registry website at \href{https://aspredicted.org/yd32-r96n.pdf}{https://aspredicted.org/yd32-r96n.pdf}.

\subsection{Supplementary Experiment: Projection onto Other People}
\label{sec:human_on_human}

As we discuss in the introduction, self-projection relates to the well-known projection bias from the behavioral economics literature. Our main experiment was designed to establish that an analogous misperception exists in the domain of beliefs about GenAI. In this section, we describe a supplementary experiment that was designed and conducted later. The goal of this experiment was to compare the magnitude of self-projection onto GenAI with projection onto other people.

The supplementary experiment was nearly identical to the main experiment, except that subjects were asked to predict the average choices of other human participants rather than those of GenAI. Specifically, the first part of this follow-up study was identical to the original experiment. Subjects were asked to make choices in the decision problems, presented in a random order. In the second part, subjects were asked to predict the average choices made by other subjects, again presenting the problems in a random order. (We did not ask subjects to make predictions about the ``dictator200'' problem since, as in the main experiment, no subject faced this problem in the first part.)
We conducted this experiment in September 2025 with 300 new subjects who were recruited on Prolific.

\section{Experimental Results}
\label{sec:main_results}

\subsection{Descriptive Statistics}
Out of 300 subjects who participated in the main experiment, 62.7\% identified as women, 35.3\% identified as men, and the rest did not provide an answer. Subjects' average age was 37 (s.d. 13). Consistent with our requirement that subjects live in the U.S., the majority of subjects were born in the U.S., with 64\% identifying as White, 14\% identifying as Black, 7.7\% identifying as Asian, and the rest identifying as mixed or as belonging to other racial groups.

\begin{figure}[h!]
    \centering
    \begin{subfigure}[t]{0.45\textwidth}
        \includegraphics[width=\textwidth]{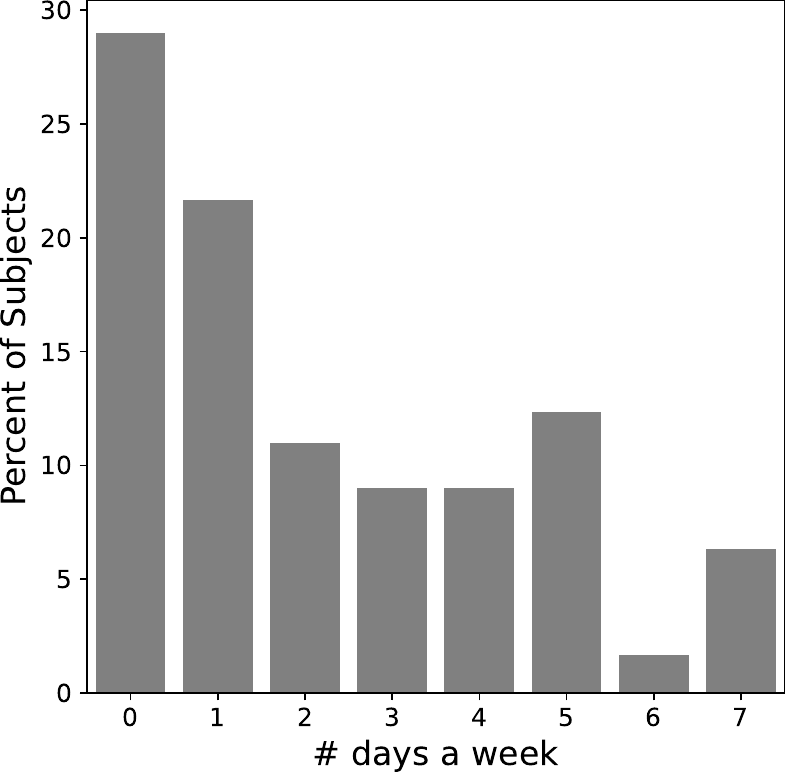}
        \caption{Intensity of GenAI Use}
    \end{subfigure}
    \hfill
    \begin{subfigure}[t]{0.45\textwidth}
        \includegraphics[width=\textwidth]{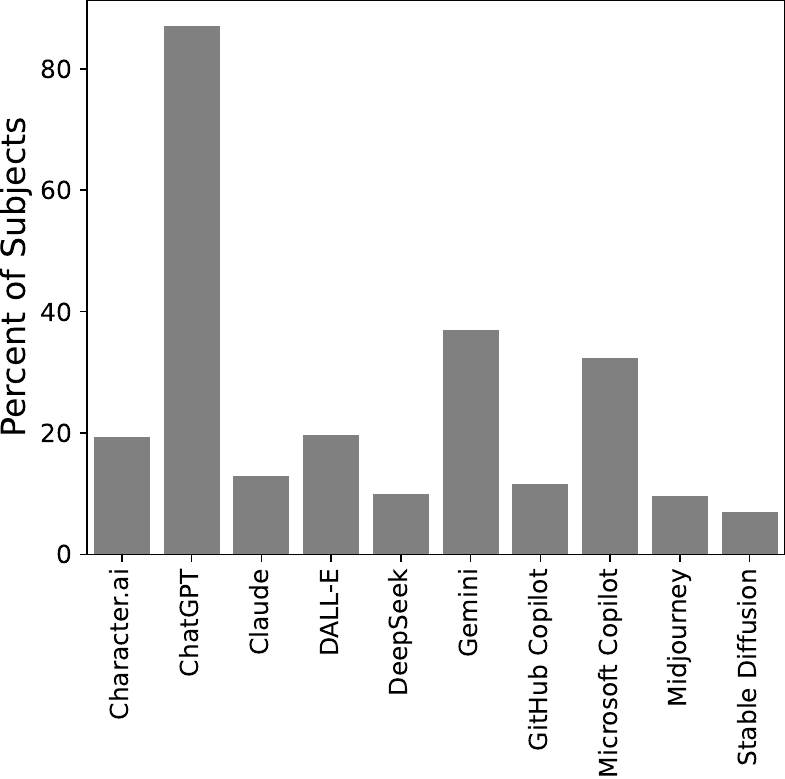}
        \caption{Experience with GenAI Models}
    \end{subfigure}

    \vspace{1em}

    \begin{subfigure}[b]{0.45\textwidth}
        \includegraphics[width=\textwidth]{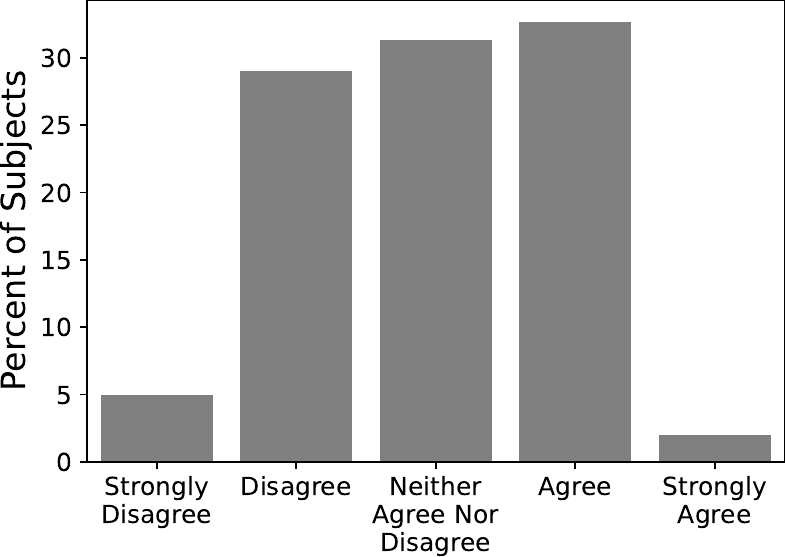}
        \caption{GenAI Makes Similar Decisions}
    \end{subfigure}
    \hfill
    \begin{subfigure}[b]{0.45\textwidth}
        \includegraphics[width=\textwidth]{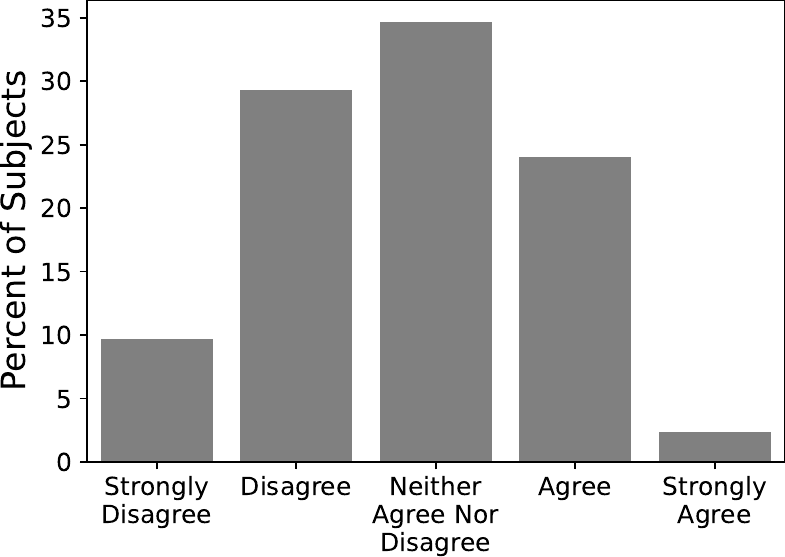}
        \caption{GenAI Makes Better Decisions}
    \end{subfigure}
    \caption{Experience and attitudes toward GenAI. (a) Distribution of responses to the question: ``In a typical week, on how many days do you use generative AI tools?'' (b) Percentage of subjects who have used various GenAI models before. (c) Degree of agreement with the statement: ``Decisions made by AI are on average similar to decisions made by humans.'' (d) Degree of agreement with the statement: ``On average, AI makes better decisions than humans.''}
    \label{fig:survey_results}
\end{figure}

\cref{fig:survey_results} summarizes the subjects' answers to the survey questions regarding their exposure, usage, and attitudes towards GenAI (administered at the end of the study). \cref{fig:survey_results}(a) shows that, on average, subjects report using GenAI two days in a typical week. \cref{fig:survey_results}(b) displays the percentages of subjects who have used various GenAI models at least once before. The survey also asked the subjects whether they agree that GenAI makes decisions similar to those of humans and whether they agree that GenAI makes  better decisions than humans.
Figures~\ref{fig:survey_results}(c) and (d) show the distributions of responses. The results reveal considerable heterogeneity in attitudes among subjects, though few hold extreme views on either statement.

\cref{tab:task_responses_summary} summarizes the distributions of subjects' choices and predictions,  along with the average choices by  GPT-4o. For ease of comparison, we also reproduce the average responses among the Amazon Mechanical Turk (MTurk) subjects in \citet{Yariv} whenever they are available.
The table shows that subjects exhibit substantial variations in their choices and predictions. Additionally, the average responses in our Prolific subject pool are  similar to those documented among MTurk users in \citet{Yariv}. Another persistent pattern is that GenAI makes more generous choices than humans in dictator games, consistent with the finding in \citet{mei2024turing}.

\begin{table}[t]
    \centering
    \caption{Subjects' Choices, Subjects' Predictions, and GenAIs Choices (Main Experiment).}
    \vspace{-.5cm}
     \[
\begin{array}{lrrrrrr}
\toprule
 & \multicolumn{2}{c}{\text{Human Choice}} & \multicolumn{2}{c}{\text{Human Prediction}} & \text{GPT-4o} & \text{Mturk} \\
 \cline{2-3}  \cline{4-5}
 & \text{Mean} & \text{Std. Dev.} & \text{Mean} & \text{Std. Dev.} & \text{Choice} & \text{SY } \\
\midrule
\text{risk100} & 31.99 & 27.68 & 36.48 & 24.37 & 11.77 & 44\\
\text{risk200} & 91.80 & 55.02 & 96.84 & 48.58 & 123.10 & 98\\
\text{discounting} & 300.48 & 73.91 & 282.26 & 69.13 & 174.85 & \text{N/A} \\
\text{dictator100} & 27.27 & 25.67 & 31.68 & 25.70 & 48.94 & 26\\
\text{dictator300} & 80.36 & 72.92 & 90.38 & 73.93 & 142.61 & 74\\
\text{dictator100x2} & 28.92 & 27.52 & 32.61 & 27.41 & 64.02 & 30\\
\text{dictator100x0.5} & 27.89 & 28.85 & 29.70 & 27.46 & 38.59 & 25 \\
\text{prisoner} & 57.33 & 49.54 & 51.17 & 25.90 & 10.08 & 43 \\
\text{beauty} & 50.65 & 23.77 & 48.57 & 19.42 & 24.13 & \text{N/A}\\
\text{dictator200} & \text{N/A} & \text{N/A} & 62.32 & 50.75 & 95.76 & \text{N/A}\\
\bottomrule
\end{array}
\]
    \label{tab:task_responses_summary}
    \parbox{\linewidth}{\small \textit{Note:} For Decision Problem 8 (``prisoner,'' the prisoner's dilemma game), we report the percentage rate of cooperation for choices and predictions.  Decision Problem 10 (``dictator200'') served only as a prediction task (and not as a  choice task). ``GPT-4o Choice'' is the average choice that GPT-4o makes when instructed to act on behalf of a human user. The exact prompt is described in \cref{sec:prompts}. The last column, titled ``Mturk SY,'' reproduces responses from \citet{Yariv} whenever they are available. While \citet{Yariv} do not report the average choice in their ``discounting''  elicitation, they report an average monthly discounting rate of $0.67$. }
\end{table}

\subsection{Anthropomorphic Projection}\label{subsec:PRA}

To assess the degree of anthropomorphic projection, we need to  compare subjects' predictions about the average GenAI choice in each problem to the actual average GenAI choice as well as the average human-subject choice.
For this purpose, we pre-registered the \emph{relative prediction accuracy} (RPA) measure, which is given by the following formula:
\begin{equation} \label{eq:RPA}
	\hat{r}_j = 1 - \frac{|\bar{P}_j-\bar{Y}_j|}{|\bar{P}_j-\bar{Y}_j|+|\bar{P}_j-\bar{X}_j|}.
\end{equation}
Here, $j$ is a task,  $\bar{P}_j$ is the subjects' average prediction, $\bar{X}_j$ is the subjects' average choice, and $\bar{Y}_j$ is the GenAI's average choice (all quantities for task $j$). An RPA of 1 occurs when the average human prediction fully matches the average GenAI choice. An RPA of 0 occurs when the prediction fully matches the average human choice. An RPA of 0.5 occurs when the average prediction is equidistant between the average GenAI choice and the average human choice.

RPA is a consistent estimator for the $r$ parameter from our model in \cref{sec:theory}. Suppose a subject of type $\theta$ chooses $a(\omega,\theta)$
and predicts GenAI's choice to be \[\rho\cdot a(\omega,\theta)+(1-\rho)\mathbb{E}_{\theta}[a(\omega,\theta)]+r\cdot b(\omega)+\varepsilon,\] where $\varepsilon$ is a mean-zero error term independent of other random variables. As the number of subjects grows large for problem $j$, we get  \[\bar{P}_j \to  \mathbb{E}_{\theta}[\rho\cdot a(\omega,\theta)+(1-\rho)\mathbb{E}_{\theta}[a(\omega,\theta)]+r\cdot b(\omega)]=\mathbb{E}_{\theta}[a(\omega,\theta)]+r\cdot b(\omega).\]
So, as the number of subjects grows, RPA estimates
\begin{align*}
  1-\frac{|\mathbb{E}_{\theta}[a(\omega,\theta)]+r\cdot b(\omega)-(\mathbb{E}_{\theta}[a(\omega,\theta)]+b(\omega))|}{|\mathbb{E}_{\theta}[a(\omega,\theta)]+r\cdot b(\omega)-(\mathbb{E}_{\theta}[a(\omega,\theta)]+b(\omega))|+|\mathbb{E}_{\theta}[a(\omega,\theta)]+r\cdot b(\omega)-(\mathbb{E}_{\theta}[a(\omega,\theta)])|},
\end{align*}
which simplifies to \[ 1-\frac{(1-r)\cdot|b(\omega)|}{(1-r)\cdot|b(\omega)|+r\cdot|b(\omega)|} = r.\]

\begin{table}[t]
    \centering
    \caption{Summary of the Main Results }

    \begin{tabular}{lcccc}
\toprule
 & RPA & 95\% CI for RPA & $\hat{\rho}_j$ & SE($\hat{\rho}_j$) \\
\midrule
risk100 & 0.154 & [0.052, 0.229] & $0.368^{***}$ & 0.059 \\
risk200 & 0.161 & [0.014, 0.328] & $0.442^{***}$ & 0.052 \\
discounting & 0.145 & [0.083, 0.203] & $0.459^{***}$ & 0.054 \\
dictator100 & 0.204 & [0.050, 0.334] & $0.347^{***}$ & 0.070 \\
dictator300 & 0.161 & [0.026, 0.291] & $0.435^{***}$& 0.065 \\
dictator100x2 & 0.105 & [0.019, 0.187] & $0.493^{***}$ & 0.061 \\
dictator100x0.5 & 0.170 & [0.010, 0.446] & $0.383^{***}$ & 0.062 \\
prisoner & 0.130 & [0.018, 0.226] & $0.149^{***}$ & 0.028 \\
beauty & 0.078 & [0.005, 0.163] & $0.401^{***}$ & 0.054 \\
\bottomrule
\end{tabular}

    \label{tab:main_result}

    \vspace{0.4cm}
    \parbox{\linewidth}{\small \textit{Note:} RPA is calculated according to the formula provided in \cref{eq:RPA}. RPA values lower than 0.5 indicate that the average prediction about GenAI's choice  is closer to the average human-subject choice than  the actual GenAI choice. 95\% CI for RPA is the bootstrap 95\% confidence interval of RPA, calculated using 10{,}000 resamples (with replacement).
      $\hat{\rho}_j$ is an estimate of $\rho_j$, a linear regression coefficient that measures how subjects' predictions about GenAI choices correlate with their own choices in the same problem
      (see \cref{eq:main}). SE($\hat{\rho}_j$) is its robust (HC1) standard error. All $\hat{\rho}_j$'s  are statistically significant at the 1\% level.}
\end{table}

Column (1) of \cref{tab:main_result} presents the RPA for each problem.\footnote{Our pre-registration specifies that if the average GenAI choice is too close to the average human-subject choice in any problem (in particular, if the two are within 0.1 standard deviations of human subjects' choices), then we will exclude the problem from the RPA analysis. This did not happen for any of the problems.} Across all problems,  RPA ranges between 0.08 and 0.20 and all 95\% confidence intervals for RPA are well below 0.5. That is, subjects' average predictions about the  GenAI choice are substantially closer to the average human-subject choice than they are to the actual average GenAI choice.  Additionally, \cref{tab:main_result_drop_copy} shows that the RPA decreases even further when we exclude the 11\% of subjects with detected copying behavior (some of whom may have queried an LLM to form their predictions).   Altogether, our findings support the hypothesis of anthropomorphic projection:  subjects, on average, overestimate the similarity between the average GenAI choice and the average human choice.

\subsection{Self-Projection}\label{subsec:self-projection}
Next, we investigate the degree to which subjects' predictions about GenAI's choices are positively correlated with their own choices in the same problem.  For each problem $j$, we  run a linear regression to estimate the following pre-registered model
\begin{equation} \label{eq:main}
	P_{ij} = \alpha_j + \rho_j \cdot X_{ij} + \varepsilon_{ij},
\end{equation}
where $P_{ij}$ is subject $i$'s prediction of GenAI's choice in problem $j$, and $X_{ij}$ is the de-meaned version of subject $i$'s own choice for problem $j$ (that is,  $X_{ij}$  is $i$'s choice minus $\bar{X}_j$, the average choice among all subjects for problem $j$). The coefficient of interest is the slope of the regression. Using the notation of our model (\cref{sec:theory}),  suppose subject $i$'s choice is $\tilde{X}_{ij}=a(\omega_j,\theta_i)$ and
subject $i$'s prediction is generated by  \[P_{ij}=\rho^\star \cdot a(\omega_j,\theta_i)+(1-\rho^\star) \cdot \mathbb{E}_{\theta}[a(\omega,\theta)]+r^\star \cdot b(\omega_j)+\varepsilon_{ij},\] where $\rho^\star$ is the self-projection coefficient for this problem and  $\varepsilon_{ij}$ is independent of other random variables. Rearranging terms, we get
\[
P_{ij}=\underset{\text{constant in } i}{\underbrace{[(1-\rho^\star) \cdot \mathbb{E}_{\theta}[a(\omega_j,\theta)]+r^\star \cdot b(\omega_j)]}}+\rho^\star \cdot \tilde{X}_{ij}+\varepsilon_{ij}.
\]
 So the regression coefficient on subjects' actions identifies $\rho^\star$.
(In our regressions, we use de-meaned versions of subjects' choices, but this does not change the slope coefficient.)

The two rightmost columns of \cref{tab:main_result} report our estimates of  $\rho_j$ (additional details are provided in \cref{tab:individual_regressions}). Across all problems, our estimates
of $\rho_j$  are positive, substantial, and statistically different from zero at the 1\% level.
These findings are consistent with subjects projecting their personal traits onto GenAI.  For example, subjects revealed to be more risk-seeking through their choices (i.e., those who wager more tokens in the two risk-domain problems) tend to also believe that GenAI will behave in a more risk-seeking way, and vice versa for the more risk-averse individuals.

One may wonder if our findings result from  subjects memorizing their choices for every problem in the first part of the experiment and simply repeating them as their predictions or using them as anchors for their predictions in the second part of the experiment.  To rule out this possibility, we analyze predictions in dictator200, a problem that was not used as a choice task in the first part of the experiment. We estimate regression models of the form
\begin{equation} \label{eq: cross predict}
	P_{ij} = \alpha_{jk} + \rho_{jk} \cdot X_{ik} + \varepsilon_{ij},
\end{equation}
where $P_{ij}$ is subject $i$'s prediction of GenAI's choice in problem $j$ and $X_{ik}$ is the de-meaned version of subject $i$'s own choice for a different problem $k$.

We set $j= \text{dictator200}$. For regressors, we separately include the subjects' choices in four other dictator problems and two risk problems (as $k$).  \cref{tab:cross_give1} presents the results.  We find that subjects' choices from the dictator problems are highly correlated with their predictions of GenAI choice in dictator200, with all coefficient estimates $\hat{\rho}_{jk}$ being positive and statistically significant at the 1\% level. This is consistent with self-projection operating through a channel where subjects project their social-preference parameter onto the GenAI, so a generous subject both chooses to give away more tokens in the four dictator-type choice tasks and predicts the GenAI would give away more tokens in the new prediction task that was previously unseen. By contrast, choices from the two  problems that belong to a different domain (risk problems) have much less explanatory power (as measured by $R^2$). Additionally, the estimated coefficient on one of the risk problems is not statistically significant at standard levels.

\begin{table}[!htbp] \centering
  \caption{Human Choices and Predictions About GenAI Choice in Related Problems}
\begin{adjustbox}{scale=0.95}
\begin{tabular}{lcccccc}
\toprule
& \multicolumn{6}{c}{Dependent variable: P\_dictator200} \\
\cmidrule(lr){2-7}
 & (1) & (2) & (3) & (4) & (5) & (6) \\
\midrule
 constant & 62.323$^{***}$ & 62.323$^{***}$ & 62.323$^{***}$ & 62.323$^{***}$ & 62.323$^{***}$ & 62.323$^{***}$ \\
& (2.651) & (2.665) & (2.777) & (2.699) & (2.882) & (2.925) \\
 X\_dictator100 & 0.849$^{***}$ & & & & & \\
& (0.137) & & & & & \\
 X\_dictator300 & & 0.292$^{***}$ & & & & \\
& & (0.048) & & & & \\
 X\_dictator100x2 & & & 0.598$^{***}$ & & & \\
& & & (0.132) & & & \\
 X\_dictator100x0.5 & & & & 0.691$^{***}$ & & \\
& & & & (0.119) & & \\
 X\_risk100 & & & & & 0.348$^{***}$ & \\
& & & & & (0.122) & \\
 X\_risk200 & & & & & & 0.078$^{}$ \\
& & & & & & (0.059) \\
\midrule
 Observations & 300 & 300 & 300 & 300 & 300 & 300 \\
 $R^2$ & 0.184 & 0.176 & 0.105 & 0.154 & 0.036 & 0.007 \\
\midrule
\multicolumn{7}{p{0.95\textwidth}}{\textit{Note:} Robust standard errors are reported in parentheses. $^{*}$p$<$0.1; $^{**}$p$<$0.05; $^{***}$p$<$0.01.} \\
\bottomrule
\end{tabular}
\end{adjustbox}
\label{tab:cross_give1} \end{table}

We extend this analysis to problems that appeared as both choice tasks and prediction tasks. In \cref{tab:cross_give2,tab:cross_give3}, we regress predictions in one dictator problem on choices in another dictator problem. In \cref{tab:cross_investment}, we regress predictions in one risk problem on choices in the other risk problem and on choices in the dictator problems.
The results show that in every case, the coefficient estimate $\hat{\rho}_{jk}$ of own choices in a related problem is positive and statistically significant at the 1\% level. Furthermore, mirroring the findings from \cref{tab:cross_give1}, choices from dictator problems have much less explanatory power (as measured by $R^2$) compared to choices from the other risk problem in explaining the subjects' predictions in risk problems.

\paragraph{Summary.} We find that, as a group, human subjects overestimate the similarity between the average human choice and the average GenAI choice. Additionally, at the individual level, human subjects overestimate the correlation between their own choices and GenAI choices in every problem. We also provide evidence that suggests that this correlation may arise from human subjects projecting their traits (such as domain-specific preference parameters) onto the AI.

\subsection{Heterogeneity Analyses}

In \cref{sec:heterogeneity}, we explore how the degree of self-projection varies with exposure to  GenAI, with attitudes toward GenAI, with time spent on each prediction task, and with gender. We find limited evidence of heterogeneity along any of these dimensions. In particular, these results suggest that increased experience with GenAI and longer deliberation time are not associated with significant reductions in self-projection.

\subsection{Projection onto Other Humans}
\label{sec:human_on_human}

Out of 300 subjects in the supplementary experiment, 54.7\% identified as women, 44.7\%
identified as men, and the rest did not provide an answer. Subjects’ average age was 44
(s.d. 13). Consistent with our requirement that subjects live in the U.S., most subjects were born in the U.S., with 72\% identifying as White, 12.7\% identifying as Black, 5\% identifying as Asian, and the rest identifying as mixed or as belonging to other racial
groups.

\begin{table}[t]
    \centering
    \caption{Comparing Self-Projection onto GenAI with Projection onto Other Humans}

    \begin{tabular}{lllrrr}
\toprule
 & $\hat{\rho}_j^{AI}$ & SE($\hat{\rho}_j^{AI}$)  & $\hat{\rho}_j^{Human}$ & SE$(\hat{\rho}_j^{Human})$ & $\hat{\rho}_j^{AI}$/$\hat{\rho}_j^{Human}$ \\
\midrule
risk100 & $0.368^{***}$ & 0.059  & $0.334^{***}$ & 0.044 & 1.100 \\
risk200 & $0.442^{***}$ & 0.052  & $0.356^{***}$ & 0.040 & 1.241 \\
discounting & $0.459^{***}$ & 0.054  & $0.568^{***}$ & 0.042 & 0.809 \\
dictator100 & $0.347^{***}$ & 0.070  & $0.551^{***}$ & 0.057 & 0.629 \\
dictator300 & $0.435^{***}$ & 0.065  & $0.579^{***}$ & 0.044 & 0.751 \\
dictator100x2 & $0.493^{***}$ & 0.061  & $0.474^{***}$ & 0.056 & 1.040 \\
dictator100x0.5 & $0.383^{***}$ & 0.062  & $0.590^{***}$ & 0.052 & 0.650 \\
prisoner & $0.149^{***}$ & 0.028  & $0.223^{***}$ & 0.023 & 0.667 \\
beauty & $0.401^{***}$ & 0.054  & $0.442^{***}$ & 0.043 & 0.908 \\
\bottomrule
\end{tabular}

    \label{tab:main_result_sept}
    \vspace{0.2cm}
    \parbox{\linewidth}{\vspace{6pt} \small \textit{Note:} All $\hat{\rho}_j$'s  are statistically significant at the 1\% level.}
\end{table}

\cref{tab:main_result_sept} presents the estimated correlation coefficients from \cref{eq:main}. For convenience, we present them (denoted $\hat{\rho}_j^{AI}$) alongside the coefficients from the main experiment (which we denote $\hat{\rho}_j^{Human}$). We also report the ratio $\hat{\rho}_j^{AI} / \hat{\rho}_j^{Human}$, which ranges from 0.63 to 1.24. In most problems, the ratio is lower than 1, suggesting higher levels of projection onto other humans.
However, the difference is not large. Averaged across decision problems, the ratio of the estimated regression coefficients is 0.87. We interpret the findings of \cref{tab:main_result_sept} as
evidence that self-projection onto GenAI is substantively similar in magnitude to projection onto other humans.

By contrast, comparing \cref{tab:task_responses_summary,tab:task_responses_summary_sept} reveals a key difference between subjects' average predictions in the main experiment and the supplementary experiments: The average human prediction of other humans' choices is substantially more accurate than the average human prediction of GenAI choices.\footnote{Since the studies were not conducted simultaneously, the participant pools differ in demographics. We therefore do not emphasize these difference.} A possible explanation is that humans are (for the time being) better at predicting the choices of other humans than they are at predicting GenAI. This possibility is also consistent with $R^2$ being higher across decision problems in the regressions that underlie \cref{tab:main_result_sept}.

\begin{table}[t]
    \centering
    \caption{Subjects' Choices and Predictions about Other People's Choices (Supplementary Experiment)}
     \begin{tabular}{lrrrrl}
\toprule
 & \multicolumn{2}{c}{Human Choice} & \multicolumn{2}{c}{Human Prediction} & Mturk \\
 & Average & Std. Dev. & Average & Std. Dev. & SY \\
\midrule
risk100 & 31.02 & 29.51 & 33.16 & 21.52 & 44 \\
risk200 & 85.11 & 58.68 & 84.02 & 41.37 & 98 \\
discounting & 296.42 & 76.24 & 290.63 & 70.29 & N/A \\
dictator100 & 20.13 & 22.53 & 23.93 & 20.43 & 26 \\
dictator300 & 59.16 & 65.81 & 67.91 & 58.39 & 74 \\
dictator100x2 & 20.66 & 23.27 & 24.52 & 20.02 & 30 \\
dictator100x0.5 & 19.53 & 24.11 & 22.76 & 22.02 & 25 \\
prisoner & 55.33 & 49.80 & 56.30 & 22.08 & 57 \\
beauty & 49.30 & 21.43 & 48.54 & 17.46 & N/A \\
\bottomrule
\end{tabular}

    \label{tab:task_responses_summary_sept}
    \vspace{0.2cm}
    \parbox{\linewidth}{\vspace{6pt} \small \textit{Note:} For Decision Problem 8 (``prisoner,'' the prisoner's dilemma game), we report the percentage rate of cooperation for choices and predictions. The last column, titled ``Mturk SY,'' reproduces responses from \citet{Yariv} whenever they are available. While \citet{Yariv} do not report the average choice in their ``discounting''  elicitation, they report an average monthly discounting rate of $0.67$.}
\end{table}

\subsection{Measurement Error Correction} If subjects do not optimize perfectly or are not fully attentive, then their responses in the choice tasks may be noisy measurements of their actual optimal choices in the decision problems.\footnote{We thank John Conlon for this suggestion.} As a result, the OLS estimates of $\rho_j$ in \cref{eq:main} may suffer from attenuation bias (namely, underestimate the extent of self-projection onto AI or projection onto other humans).

To address this issue---which could theoretically affect the interpretation of the results---we focus on the risk and dictator domains, which consist of multiple related decision problem. We estimate two-stage least squares (2SLS) regressions, using choices in related tasks as instruments. For each of the two risk problems, we use the other risk problem as an instrument, and for each dictator problem, we use the other three dictator problems as instruments.

\cref{tab:iv_result} reports the results. Indeed, across both experiments, the IV estimates are uniformly larger than the OLS estimates, consistent with attenuation bias in the OLS regressions. We note that all first-stage $F$-statistics are well above conventional thresholds. Additionally, since dictator problems are over-identified, we also report Hansen~$J$-test $p$-values. With one exception, we cannot reject the validity of our instruments.

\begin{table}[t]
\centering
\caption{IV/2SLS Estimates of Self-Projection Coefficients}\label{tab:iv_result}
\begin{tabular}{lcccccc}
\toprule
 & \multicolumn{2}{c}{OLS} & \multicolumn{2}{c}{IV/2SLS} & First-stage & Hansen $J$ \\
\cmidrule(lr){2-3} \cmidrule(lr){4-5}
Task & $\hat{\rho}$ & SE & $\hat{\rho}$ & SE & $F$-stat & $p$-value \\
\midrule
\multicolumn{7}{l}{\textit{Main Experiment: Predictions about GenAI Choices}} \\[2pt]
risk100 & $0.368^{***}$ & 0.059 & $0.492^{***}$ & 0.094 & 103.4 & --- \\
risk200 & $0.442^{***}$ & 0.052 & $0.500^{***}$ & 0.081 & 198.6 & --- \\
dictator100 & $0.347^{***}$ & 0.070 & $0.414^{***}$ & 0.076 & 471.7 & 0.504 \\
dictator300 & $0.435^{***}$ & 0.065 & $0.479^{***}$ & 0.079 & 308.6 & 0.747 \\
dictator100x2 & $0.493^{***}$ & 0.061 & $0.545^{***}$ & 0.085 & 89.4 & 0.010 \\
dictator100x0.5 & $0.383^{***}$ & 0.062 & $0.482^{***}$ & 0.075 & 345.3 & 0.493 \\
\midrule
\multicolumn{7}{l}{\textit{Supplementary Experiment: Predictions about Other People's Choices}} \\[2pt]
risk100 & $0.334^{***}$ & 0.044 & $0.403^{***}$ & 0.081 & 66.0 & --- \\
risk200 & $0.356^{***}$ & 0.040 & $0.430^{***}$ & 0.078 & 83.2 & --- \\
dictator100 & $0.551^{***}$ & 0.057 & $0.676^{***}$ & 0.052 & 337.2 & 0.347 \\
dictator300 & $0.579^{***}$ & 0.044 & $0.688^{***}$ & 0.053 & 173.4 & 0.103 \\
dictator100x2 & $0.474^{***}$ & 0.056 & $0.639^{***}$ & 0.071 & 61.9 & 0.091 \\
dictator100x0.5 & $0.590^{***}$ & 0.052 & $0.623^{***}$ & 0.057 & 123.7 & 0.405 \\
\midrule
\multicolumn{7}{p{0.95\textwidth}}{\textit{Note:} $^{***}$p$<$0.01. IV/2SLS uses choices in related tasks as instruments for the endogenous regressor $X_{ij}$. Standard errors are heteroskedasticity-robust (HC1), with second-stage SEs corrected using residuals from actual $X$. Hansen $J$-test $p$-values are reported for overidentified specifications; ``---'' indicates just-identified.} \\

\end{tabular}
\end{table}

\subsection{Interpretation}
In this section, we interpret our experimental findings in light of our model. To this end, we recall the parameter restrictions posed by different special cases of the model (\cref{sec:theory}).

First, as the experiment strongly rejects $\rho=0$ (\cref{tab:iv_result}), any parametrization of the model that precludes self-projection is ruled out. This includes  ``\emph{correctly specified beliefs},'' and ``\emph{no self-projection, some anthropomorphic projection}.''

Second, as the experiment strongly rejects $r=1$ (\cref{tab:main_result}), any parametrization of the model  that precludes anthropomorphic projection  is ruled out. This includes ``\emph{no anthropomorphic projection, some self-projection}'' (in addition to correctly specified beliefs)

Third, we show that the experiment rejects the parametric restriction $r+\rho=1$.
To do so, in \cref{tab:rpa_plus_rho_combined} we use bootstrap resampling to calculate a 95\% confidence interval for the sum $\text{RPA}_j + \rho_j $ in every problem $j$. The confidence intervals lie completely below 1 in every problem. This result holds even if we use the (higher) 2SLS estimators for $\rho_j$. Consequently, models where beliefs take the form of a``\emph{weighted average of self and GenAI}'' are ruled out.

In sum, the experimental evidence suggest that anthropomorphic projection and self-projection are two distinct phenomena that co-exist in our subject population.

\begin{table}[t]
\centering
\caption{95\% Confidence Intervals for RPA$_j$ + $\hat{\rho}_j$}
\label{tab:rpa_plus_rho_combined}
\begin{tabular}{lcccc}
\toprule
 & \multicolumn{2}{c}{OLS} & \multicolumn{2}{c}{IV/2SLS} \\
\cmidrule(lr){2-3} \cmidrule(lr){4-5}
Task & RPA$_j$ + $\hat{\rho}_j$ & 95\% CI & RPA$_j$ + $\hat{\rho}_j^{IV}$ & 95\% CI \\
\midrule
risk100 & 0.522 & [0.355, 0.673] & 0.646 & [0.420, 0.879] \\
risk200 & 0.603 & [0.410, 0.809] & 0.661 & [0.434, 0.891] \\
discounting & 0.604 & [0.482, 0.721] & --- & --- \\
dictator100 & 0.550 & [0.330, 0.759] & 0.618 & [0.398, 0.824] \\
dictator300 & 0.596 & [0.394, 0.795] & 0.640 & [0.422, 0.855] \\
dictator100x2 & 0.598 & [0.439, 0.755] & 0.650 & [0.445, 0.829] \\
dictator100x0.5 & 0.553 & [0.330, 0.883] & 0.652 & [0.410, 0.982] \\
prisoner & 0.279 & [0.160, 0.385] & --- & --- \\
beauty & 0.479 & [0.350, 0.611] & --- & --- \\
\bottomrule
\end{tabular}
\vspace{0.2cm}
    \parbox{\linewidth}{\vspace{6pt} \small \textit{Note:} The 95\% CI are the bootstrap 95\% confidence interval of $RPA_j+\rho_j$, calculated using 10{,}000 resamples (with replacement). Specifically, we do not rely on the RPA bootstrap confidence intervals and do not make any assumptions about the correlation between estimates of RPA and $\rho$.}
\end{table}

\section{Concluding Discussion}

This paper provides evidence  that people overestimate the degree to which GenAI choices are aligned with human preferences in general (anthropomorphic projection) and with their own preferences in particular (self-projection). We find limited evidence that experience attenuates these misperceptions. We show theoretically that these misperceptions lead to over-delegation to GenAI and interact with the true degree of AI alignment to produce complex welfare implications.

We are not the first to study selective delegation to AI. We view the main contribution of our work as documenting the individual-level phenomenon of self-projection. This  is facilitated by our focus on  economic decision environments that involve trade-offs, where agents' optimal actions depend on their preferences.

Our findings raise many interesting questions. For example, how can we  debias self-projection?\footnote{\citet{dreyfuss2024human} report on measures for regulating the degree of anthropomorphic projection in problems that involve factual answers and not economic trade-offs.}  What are conducive design principles for GenAI agents in light of users who exhibit self-projection and anthropomorphic projection? (Specifically, should GenAI sometimes defer to the user, similar to \citet{Noti}?) Will self-projection persist in the long run? We leave these exciting questions for future research.

\appendix
\setcounter{figure}{0}
\renewcommand{\thefigure}{A.\arabic{figure}}

\setcounter{table}{0}
\renewcommand{\thetable}{A.\arabic{table}}

\section{Proofs}

\subsection{Proof of Proposition \ref{prop:group_level}}

\begin{proof}

The agent expects a utility of $-r^{2}\mathbb{E}[b(\omega)^{2}]$
from delegating to GenAI, $-c$ from paying attention and taking the
optimal action after learning $\omega$, and $-\sigma_{\omega}^{2}$
from not paying attention and taking the ex-ante optimal action $a=0$.
In the case where $\mathbb{E}[b(\omega)^{2}]>\sigma_{\omega}^{2}$,
the expected payoff from choosing $a=0$ is strictly higher than that
of delegation, so a correctly specified agent never delegates. An agent with
$r>\frac{\sigma_{\omega}}{\sqrt{\mathbb{E}[b(\omega)^{2}]}}$ also
perceives the utility of delegation to be strictly lower than that
of choosing $a=0$, so they also never delegate. An agent with $r<\frac{\sigma_{\omega}}{\sqrt{\mathbb{E}[b(\omega)^{2}]}}$
perceives the utility of delegation to be strictly higher than that
of choosing $a=0$, so they will choose to delegate if the attention
cost is higher than $r^{2}\mathbb{E}[b(\omega)^{2}]$.

In the case where $\mathbb{E}[b(\omega)^{2}]<\sigma_{\omega}^{2}$,
both the correctly specified agent and the biased agent never choose to take
the action $a=0.$ They choose between delegating to GenAI or paying
the attention cost $c,$ depending on whether $c$ is lower than their
perceived loss from delegation, $r^{2}\mathbb{E}[b(\omega)^{2}]$.
\end{proof}

\subsection{Proof of Proposition \ref{prop:self_basic}}

\begin{proof}
An agent with type $\theta$ expects a utility of $-(1-\rho)^{2}\theta^{2}$
from delegating to GenAI, $-c$ from paying attention and taking the
optimal action after learning $\omega$, and $-\sigma_{\omega}^{2}$
from not paying attention and taking the ex-ante optimal action $a=0$.
A correctly specified agent with type $|\theta|>\sigma_{\omega}$ never delegates,
and for $|\theta|<\sigma_{\omega}$ the agent either delegates or
pays the attention cost depending on if $c$ is larger than $\theta^{2}$.
The biased agent does not delegate if $|\theta|>\sigma_{\omega}/(1-\rho)$.
For $|\theta|<\sigma_{\omega}/(1-\rho)$, the agent either delegates
or pays the attention cost depending on if $c$ is larger than $(1-\rho)^{2}\theta^{2}$.
Thus the biased agent sometimes delegates while the correctly specified agent
never delegates for $|\theta|\in(\sigma_{\omega},\sigma_{\omega}/(1-\rho))$,
and the biased agent delegates with strictly higher probability than
the correctly specified agent for $|\theta|<\sigma_{\omega}/(1-\rho).$
\end{proof}

\subsection{Proof of Proposition \ref{prop:monotonic}}

\begin{proof}
Consider two correctly specified agents with types $0\le\theta_{1}<\theta_{2}$
(other cases are symmetric). For any optimal strategy $\sigma_{2}(c)$
of agent $\theta_{2}$ that maps the cost realization to a decision
between delegation, paying attention, or taking an action without
paying attention, consider the strategy $\sigma_{1}(c)$ of $\theta_{1}$
which (i) pays attention for every $c$ where $\sigma_{2}(c)$ pays
attention; (ii) chooses $\theta_{1}$ without paying attention for
every $c$ where $\sigma_{2}(c)$ chooses $\theta_{2}$ without paying
attention; (iii) delegates to the GenAI for every $c$ where $\sigma_{2}(c)$
delegates to the GenAI. Note that $\theta_{1}$ and $\theta_{2}$ get
the same payoff if they both pay attention, and they get the same
payoff of $-\sigma_{\omega}^{2}$ when they choose actions equal to
their types without paying attention. Delegation to GenAI has an expected
payoff of $-(\theta_{1})^{2}$ for type $\theta_{1}$ and $-(\theta_{2})^{2}$
for type $\theta_{2}$, so the former is higher. This shows $\theta_{1}$'s
welfare under the optimal strategy must be weakly higher than that
of $\theta_{2},$ so welfare is monotonically decreasing in $|\theta|.$

Now consider agents who suffer from self-projection
with $\rho\in(0,1)$. All types to the right of $\sigma_{\omega}/(1-\rho)$
behave rationally. A type slightly to the left of $\sigma_{\omega}/(1-\rho)$
delegates to the GenAI when $c$ is higher than about $\sigma_{\omega}^{2}$,
but the true expected welfare from delegation is around $-\sigma_{\omega}^{2}/(1-\rho)^{2}$
whereas the true expected welfare from taking the the default action
$a=\theta$ is around $-\sigma_{\omega}^{2}$. Therefore the biased
agent with type slightly to the left of $\sigma_{\omega}/(1-\rho)$
has welfare that is discretely lower than that of the correctly specified agent
of the same type. Since the correctly specified agent's payoff is continuous
in type, this means there must be an upward jump in welfare at $\sigma_{\omega}/(1-\rho)$.
\end{proof}

\bibliographystyle{ecta}
\bibliography{bib}
\newpage
\begin{center}
\begin{LARGE}

\textbf{Online Appendix}

\end{LARGE}
\end{center}

\section{Heterogeneity Analyses}
\label{sec:heterogeneity}
In this appendix, we explore how the degree of self-projection varies along several dimensions: experience with GenAI, attitudes toward GenAI, attention (as proxied by the amount of time spent on prediction tasks), and gender.  We find limited evidence of heterogeneity along any of these dimensions.

\subsection{Experience with GenAI and Attitudes Toward GenAI}
\label{subsec:interact_GenAI_exp}

As we discuss in the introduction, some of the possible explanations for self-projection suggest that it will be attenuated as people gain more experience with GenAI. Additionally, self-projection may also affect, and be affected by, people's beliefs about the quality of GenAI decision-making. This motivates us to assess the heterogeneity of our findings with respect to experience with GenAI and attitudes toward GenAI.

To this end, we split subjects to two groups based on their survey responses and  estimate regressions of the following form:
\begin{equation} \label{eq:hetero}
	P_{ij}= \alpha_j + \rho_{j} \cdot  X_{ij} +  \delta_j \cdot G_i + \gamma_j \cdot  G_i \cdot X_{ij} + \varepsilon_{ij}
\end{equation}
where $G_i$ is an indicator variable for whether subject $i$ belongs to one of the groups.
The coefficient of interest is
$\gamma_j$. It measures if group membership is associated with lower (if $\gamma_j < 0$) or higher (if $\gamma_j > 0$) levels of self-projection in problem $j$.

We estimate the regression model of \cref{eq:hetero} using each of the following group classifications to define $G_i$:
\begin{enumerate}
	\item \texttt{Heavy User}: Subjects who reported using GenAI at least two days in a typical week (i.e., their answer was above the median).
	\item \texttt{Text-Based LLM User}: Subjects who reported  that they have used ChatGPT, Gemini, Claude, or DeepSeek before.\footnote{This group excludes subjects who have only used AI image generators like Midjourney, but not LLMs that primarily output free-form text.}
	\item \texttt{Paid User}: Subject who reported having a paid subscription to a GenAI model or application.
	\item \texttt{Agree AI Similar}: Subject who agreed or strongly agreed with the statement: \textit{``Decisions made by GenAI are on average similar to decisions made by human.''}
	\item \texttt{Agree AI Better}: Subject who agreed or strongly agreed with the statement: \textit{``On average, GenAI makes better decisions than humans.''}
\end{enumerate}

\begin{figure}[t]
    \centering
    \includegraphics[width=0.95\linewidth]{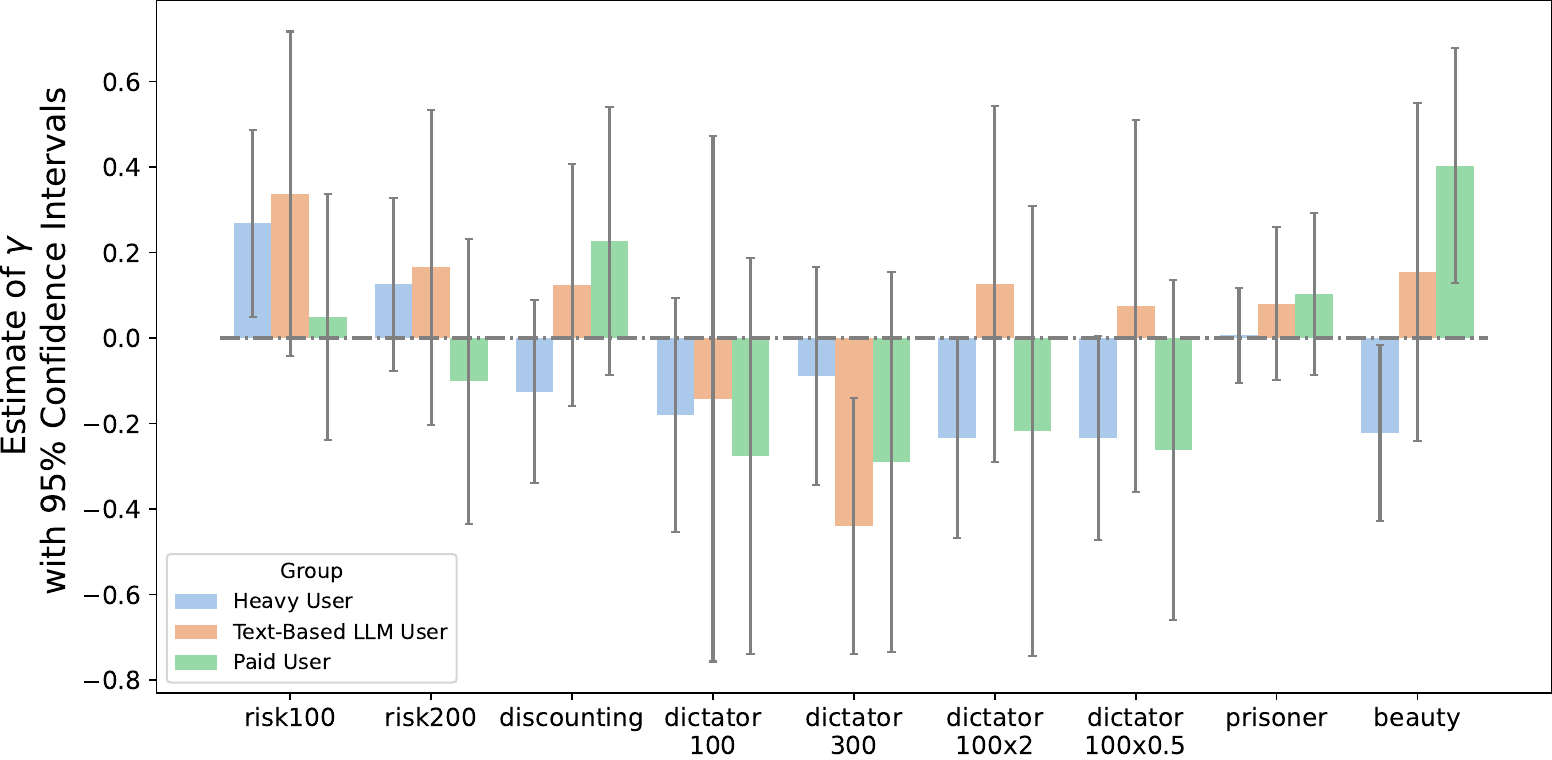}
    \caption{Heterogeneity: Experience with GenAI}
    \label{fig:group_experience}
\end{figure}

\begin{figure}[ht]
    \centering
    \includegraphics[width=0.95\linewidth]{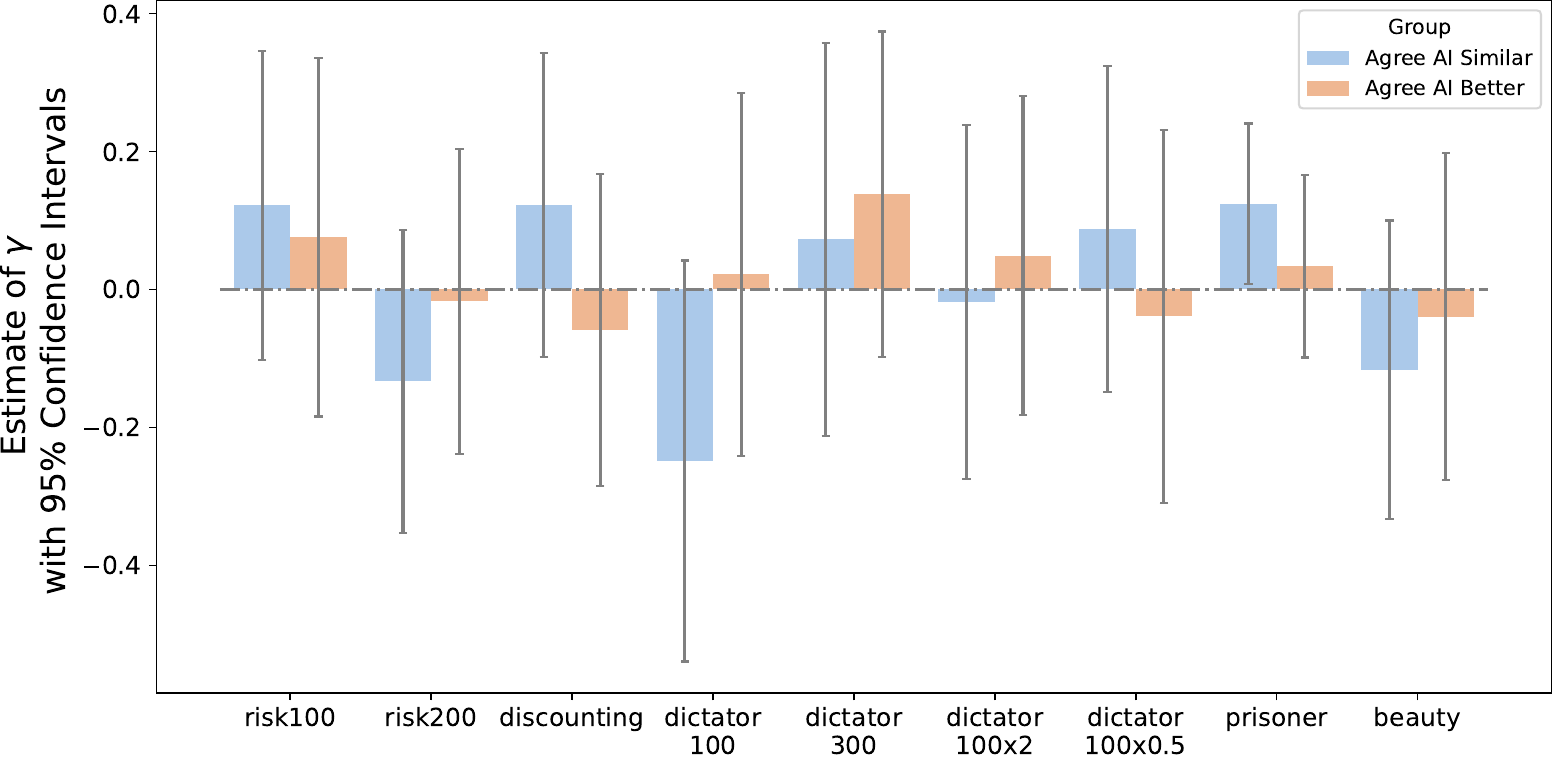}
    \caption{Heterogeneity: Attitudes Toward GenAI}
    \label{fig:group_attitude}
\end{figure}

 \cref{fig:group_experience,fig:group_attitude} plot the point estimates of $\gamma_j$ and the 95\% confidence interval for each problem  $j$  for each of the five group classifications (the full regression results are in \cref{Additional: experience}). The figures reveal that the point estimates are mixed and noisy.

To increase power, we also pool all problems together and estimate a model that imposes the assumption that $\gamma_j$ is constant across problems (recall from \cref{tab:main_result} that the magnitude of the coefficient estimate $\hat{\rho}_j$ was similar across most tasks). Specifically, we use all data to jointly estimate the following linear regression model for all problems $j$:
\begin{equation} \label{eq:hetero pooled}
	P_{i j}= \alpha_j +  \rho_{j} \cdot  X_{ij} +  \delta_j \cdot G_i + \gamma \cdot  G_i \cdot X_{ij} + \varepsilon_{ij}
\end{equation}
with standard errors  clustered at the problem level.\footnote{For this analysis, we rescale $P$ and $X$ in every problem to the range $[0,100]$ (before we demean $X$) to avoid over-weighting problems that involve larger ranges of values. Specifically, for ``discounting,'' we apply the transformation $(\text{response}-150)\times \frac{100}{400-150}$; for ``risk200,'' we divide the response by 2; and for ``dictator300,'' we divide the response by 3.\label{fn:normalization}}

\begin{table}[!htbp] \centering
  \caption{Human Choices and Predictions About GenAI Choice in the Same Problem: Heterogeneity by Exposure and Attitudes (Pooled)}
\begin{adjustbox}{scale=0.8}
\begin{tabular}{lccccc}
\toprule
& \multicolumn{5}{c}{Dependent variable: Prediction} \\
\cmidrule(lr){2-6}
 & \multicolumn{1}{c}{Heavy User} & \multicolumn{1}{c}{Text-Based LLM User} & \multicolumn{1}{c}{Paid User} & \multicolumn{1}{c}{Agree AI Similar} & \multicolumn{1}{c}{Agree AI Better}  \\
 & (1) & (2) & (3) & (4) & (5) \\
\midrule
 X$\times$Heavy User & -0.056$^{}$ & & & & \\
& (0.052) & & & & \\
 X$\times$Text-Based LLM User & & 0.082$^{**}$ & & & \\
& & (0.040) & & & \\
 X$\times$Paid User & & & -0.001$^{}$ & & \\
& & & (0.072) & & \\
 X$\times$Agree AI Similar & & & & 0.033$^{}$ & \\
& & & & (0.045) & \\
 X$\times$Agree AI Better & & & & & 0.019$^{}$ \\
& & & & & (0.017) \\
 Problem FE & Yes & Yes & Yes & Yes & Yes \\
 X$\times$Problem FE & Yes & Yes & Yes & Yes & Yes \\
 G$\times$Problem FE & Yes & Yes & Yes & Yes & Yes \\
\midrule
 Observations & 2700 & 2700 & 2700 & 2700 & 2700 \\
\midrule
\multicolumn{6}{p{0.95\textwidth}}{\textit{Note:} Standard errors are clustered at the problem level. $^{*}$p$<$0.1; $^{**}$p$<$0.05; $^{***}$p$<$0.01.} \\
\bottomrule
\end{tabular}
\end{adjustbox}
\label{tab:group} \end{table}

Estimates of $\gamma$ are reported in \cref{tab:group}. Three out of the five point estimates are positive, and one is essentially equal to zero. The only negative point estimate comes from \texttt{Heavy User}.  It is not statistically significant, and the lower  bound on the 95\% confidence interval of this estimate is $-0.16$ (to contextualize, the median estimate of $\rho_j$ across different problems is around $0.4$).

\subsubsection{Individual Accuracy}
We also investigate if experience with GenAI correlates with more accurate predictions. To measure prediction accuracy at the individual level, we compute each subject's mean normalized absolute error (MAE), comparing  their predictions  with the GenAI's choices across all problems.\footnote{We use the same normalization as described in \cref{fn:normalization} to avoid over-weighting problems that involve larger ranges of values.} In this analysis, we compare subjects' predictions to  the choices of four different LLMs.

We estimate the following linear regression model:
\begin{equation*}
\begin{split}
	MAE_{im} &= \alpha_m + \delta_{1m} \cdot GenAI\_exposure_i
	+ \delta_{2m} \cdot agree\_AI\_similar_i \\
    &+  \delta_{3m} \cdot agree\_AI\_better_i + \delta_{4m} \cdot copier_i +  \varepsilon_{im},
\end{split}
\end{equation*}
where $i$ is a subject, $m$ is a GenAI model, and $copier_i$ is an indicator for whether we detected subject $i$ copying text from the website during the experiment. Finally, $GenAI\_exposure_i$ is a dummy variable indicating ``experience.'' We separately use three measures for this indicator: 1) \texttt{Heavy User}; 2) \texttt{Paid User}; and 3) $\texttt{Model User}_{im}$, which is an indicator for subject $i$ reporting having experience with model $m$.

\cref{tab:mae_gpt-4o} presents our results with respect to GPT-4o. The coefficients of all experience measures are close to zero,  not statistically different from zero,  and  precisely estimated. Detected copying behavior is associated with a decrease of approximately  15\% in the MAE (statistically significant at the 1\% level across specifications). Finally, agreeing that AI  makes similar decisions to humans is associated  with an approximately 8\% decrease in the MAE (statistically significant at the 5\% level across specifications), while the coefficient on agreeing that AI makes better decisions is small and not statistically significant.  In \cref{tab:mae_gpt-4o-mini,tab:mae_gemini-1.5-pro,tab:mae_gemini-1.5-flash}, we find similar results for the other larger GenAI model we study (Gemini-1.5-Pro), and weaker correlations for the two smaller models.

\subsection{Self-Projection and Response Time} \label{subsec:response time}

As response times are sometimes used to measure  attention or deliberation \citep[e.g.,][]{caplin2016measuring}, we also analyze heterogeneity along the lines of slow and fast response times. To this end, we follow a similar approach to  \cref{eq:hetero pooled} and estimate  the following linear regression model using all data:

\begin{equation*}
	P_{ij}= \alpha_j +   \rho_{j} \cdot  X_{ij} + \delta_j \cdot T_{ij} + \gamma \cdot  T_{ij} \cdot X_{ij} + \varepsilon_{ij}
\end{equation*}
Here, $T_{i j}$ is an indicator for subject $i$ having spent longer than the \emph{median} time in the prediction task for problem $j$. We take two approaches for defining the median:
\begin{enumerate}
    \item \texttt{Problem Median}: Subject $i$ spent longer than the median response time across all subjects for problem $j$.
    \item \texttt{Personal Median}: Subject $i$ spent more time on prediction task $j$ than their personal median time across all prediction tasks.
\end{enumerate}
We cluster standard errors at the problem level.

\cref{tab:time} displays our results. We find that having a response time  above the problem median is associated with slightly lower levels of self-projection (point estimate $0.06$, s.e. $0.02$, $p<0.05$). The point estimate for personal median is similar, but the estimator is more noisy and is not statistically significant at standard levels.

\subsection{Self-Projection and Gender} \label{subsec: gender}
Next, we ask if members of different demographic groups display different degrees of self-projection  \citep[such a finding would have potential equity consequences, see][]{liang2022algorithmic}.
 We estimate the regression from  \cref{eq:hetero pooled}, using the group indicator $G_i$ to refer to whether the subject $i$ self-identified as female.   \cref{tab:gender} presents our results. Our estimate of $\gamma$ is close to zero and precisely estimated, suggesting limited heterogeneity along the dimension of gender.

\section{Relative Prediction Accuracy with Different GenAI Models}

In the main analysis, we followed our pre-registered plan and used  GPT-4o as the benchmark GenAI model. In this appendix, we replicate our analysis of  subjects' relative prediction accuracy  using other LLMs. Specifically, \cref{tab:llm_avg} presents the average GenAI choice for each problem and \cref{tab:llm_RPA} presents the corresponding RPA.

In most cases, the RPA is well below 0.5, but in a few cases the predictions align more closely with GenAI choices than with human subjects' choices. Comparing the smaller models (GPT-4o-mini and Gemini-1.5-Flash) with the larger models (GPT-4o and Gemini-1.5-Pro), we observe that human predictions are more aligned with the choices made by smaller models.

\begin{table}[ht]
    \centering
    \caption{GenAI Average Choices}
    \begin{tabular}{lrrrr}
\toprule 
 & GPT-4o & GPT-4o-mini & Gemini-1.5-Pro & Gemini-1.5-Flash \\
\midrule
risk100 & 11.77 & 21.42 & 100.00 & 31.59 \\
risk200 & 123.10 & 99.94 & 119.17 & 100.00 \\
discounting & 174.85 & 224.38 & 150.02 & 165.00 \\
dictator100 & 48.94 & 31.39 & 88.86 & 0.00 \\
dictator300 & 142.61 & 122.49 & 150.00 & 25.65 \\
dictator100x2 & 64.02 & 22.89 & 100.00 & 86.80 \\
dictator100x0.5 & 38.59 & 27.93 & 100.00 & 17.60 \\
prisoner & 10.08 & 93.17 & 0.00 & 50.30 \\
beauty & 24.13 & 33.84 & 22.00 & 33.00 \\
dictator200 & 95.76 & 27.36 & 100.00 & 100.00 \\
\bottomrule
\end{tabular}

    \label{tab:llm_avg}
\end{table}

\begin{table}[ht]
    \centering
    \caption{GenAI RPA}
    \begin{tabular}{lrrrr}
\toprule
 & GPT-4o & GPT-4o-mini & Gemini-1.5-Pro & Gemini-1.5-Flash \\
\midrule
risk100 & 0.154 & 0.230 & 0.066 & 0.479 \\
risk200 & 0.161 & 0.620 & 0.184 & 0.615 \\
discounting & 0.145 & 0.239 & 0.121 & 0.135 \\
dictator100 & 0.204 & 0.938 & 0.072 & 0.122 \\
dictator300 & 0.161 & 0.238 & 0.144 & 0.134 \\
dictator100x2 & 0.105 & 0.275 & 0.052 & 0.064 \\
dictator100x0.5 & 0.170 & 0.506 & 0.025 & 0.131 \\
prisoner & 0.130 & 0.128 & 0.107 & 0.876 \\
beauty & 0.078 & 0.123 & 0.072 & 0.117 \\
\bottomrule
\end{tabular}

    \label{tab:llm_RPA}
\end{table}

\section{Main Analysis Excluding Subjects with Detected Copying Behavior}

In this section, we replicate the main analyses excluding the 33 subjects (11\%) who were detected copying text at least once during the experiment. The results are summarized in \cref{tab:main_result_drop_copy}. Our measure of anthropomorphic projection---the RPAs---are slightly lower, while the $\hat{\rho}_j$ coefficients have hardly changed.

\begin{table}[h]
    \centering
    \caption{Main Results Excluding Subjects with Detected Copying Behavior}
    \begin{tabular}{lcccc}
\toprule
 & RPA & 95\% CI for RPA & $\hat{\rho}_j$ & SE($\hat{\rho}_j$) \\
\midrule
risk100 & 0.137 & [0.023, 0.223] & $0.362^{***}$ & 0.064 \\
risk200 & 0.079 & [0.005, 0.248] & $0.452^{***}$ & 0.052 \\
discounting & 0.137 & [0.075, 0.197] & $0.476^{***}$ & 0.060 \\
dictator100 & 0.170 & [0.020, 0.313] & $0.346^{***}$ & 0.074 \\
dictator300 & 0.109 & [0.007, 0.247] & $0.443^{***}$ & 0.068 \\
dictator100x2 & 0.073 & [0.005, 0.160] & $0.500^{***}$ & 0.065 \\
dictator100x0.5 & 0.094 & [0.008, 0.416] & $0.391^{***}$ & 0.064 \\
prisoner & 0.113 & [0.011, 0.212] & $0.167^{***}$ & 0.030 \\
beauty & 0.054 & [0.003, 0.137] & $0.400^{***}$ & 0.057 \\
\bottomrule
\end{tabular}

    \label{tab:main_result_drop_copy}
    \vspace{0.2cm}
    \parbox{\linewidth}{\vspace{6pt} \small \textit{Note:} RPA is calculated according to the formula provided in \cref{eq:RPA}. 95\% CI for RPA is the bootstrap 95\% confidence interval of RPA, calculated using 10{,}000 resamples (with replacement).
      $\hat{\rho}_j$ is an estimate of $\rho_j$, a linear regression coefficient that measures how subjects' predictions about GenAI choices correlate with their own choices in the same problem
      (see \cref{eq:main}). SE($\hat{\rho}_j$) is its robust (HC1) standard error. All $\hat{\rho}_j$'s are statistically significant at the 1\% level.}
\end{table}

\begin{landscape}
\section{Additional Materials}

\subsection{Additional Tables for \cref{subsec:self-projection}}

\begin{table}[!htbp] \centering
  \caption{Human Choices and Predictions About GenAI Choice in the Same Problem}
\begin{adjustbox}{scale=0.85}
\begin{tabular}{lccccccccc}
\toprule
 & \multicolumn{1}{c}{P\_risk100} & \multicolumn{1}{c}{P\_risk200} & \multicolumn{1}{c}{P\_discounting} & \multicolumn{1}{c}{P\_dictator100} & \multicolumn{1}{c}{P\_dictator300} & \multicolumn{1}{c}{P\_dictator100x2} & \multicolumn{1}{c}{P\_dictator100x0.5} & \multicolumn{1}{c}{P\_prisoner} & \multicolumn{1}{c}{P\_beauty}  \\
 & (1) & (2) & (3) & (4) & (5) & (6) & (7) & (8) & (9) \\
\midrule
 const & 36.483$^{***}$ & 96.843$^{***}$ & 282.257$^{***}$ & 31.683$^{***}$ & 90.383$^{***}$ & 32.607$^{***}$ & 29.703$^{***}$ & 51.173$^{***}$ & 48.573$^{***}$ \\
& (1.280) & (2.432) & (3.484) & (1.394) & (3.862) & (1.377) & (1.454) & (1.436) & (0.979) \\
 X & 0.368$^{***}$ & 0.442$^{***}$ & 0.459$^{***}$ & 0.347$^{***}$ & 0.435$^{***}$ & 0.493$^{***}$ & 0.383$^{***}$ & 0.149$^{***}$ & 0.401$^{***}$ \\
& (0.059) & (0.052) & (0.054) & (0.070) & (0.065) & (0.061) & (0.062) & (0.028) & (0.054) \\
\midrule
 Observations & 300 & 300 & 300 & 300 & 300 & 300 & 300 & 300 & 300 \\
 $R^2$ & 0.175 & 0.251 & 0.241 & 0.120 & 0.184 & 0.245 & 0.162 & 0.081 & 0.240 \\
\midrule
\multicolumn{10}{p{0.95\textwidth}}{\textit{Note:} Robust standard errors are reported in parentheses. $^{*}$p$<$0.1; $^{**}$p$<$0.05; $^{***}$p$<$0.01.} \\
\bottomrule
\end{tabular}
\end{adjustbox}
\label{tab:individual_regressions} \end{table}
\end{landscape}

\begin{table}[!htbp] \centering
  \caption{Human Choices and Predictions About GenAI Choice in Related Problems}
\begin{tabular}{lcccccc}
\toprule
 & \multicolumn{3}{c}{P\_dictator100} & \multicolumn{3}{c}{P\_dictator300}  \\
\cmidrule(lr){2-4} \cmidrule(lr){5-7}
 & (1) & (2) & (3) & (4) & (5) & (6) \\
\midrule
 const & 31.683$^{***}$ & 31.683$^{***}$ & 31.683$^{***}$ & 90.383$^{***}$ & 90.383$^{***}$ & 90.383$^{***}$ \\
& (1.381) & (1.429) & (1.411) & (3.913) & (4.037) & (4.008) \\
 X\_dictator100 & & & & 1.160$^{***}$ & & \\
& & & & (0.199) & & \\
 X\_dictator300 & 0.130$^{***}$ & & & & & \\
& (0.022) & & & & & \\
 X\_dictator100x2 & & 0.257$^{***}$ & & & 0.884$^{***}$ & \\
& & (0.066) & & & (0.191) & \\
 X\_dictator100x0.5 & & & 0.279$^{***}$ & & & 0.893$^{***}$ \\
& & & (0.061) & & & (0.172) \\
\midrule
 Observations & 300 & 300 & 300 & 300 & 300 & 300 \\
 $R^2$ & 0.137 & 0.076 & 0.098 & 0.162 & 0.108 & 0.121 \\
\midrule
\multicolumn{7}{p{0.95\textwidth}}{\textit{Note:} Robust standard errors are reported in parentheses. $^{*}$p$<$0.1; $^{**}$p$<$0.05; $^{***}$p$<$0.01.} \\
\bottomrule
\end{tabular}
\label{tab:cross_give2} \end{table}

\begin{table}[!htbp] \centering
  \caption{Human Choices and Predictions About GenAI Choice in Related Problems}
\begin{tabular}{lcccccc}
\toprule
 & \multicolumn{3}{c}{P\_dictator100x2} & \multicolumn{3}{c}{P\_dictator100x0.5}  \\
\cmidrule(lr){2-4} \cmidrule(lr){5-7}
 & (1) & (2) & (3) & (4) & (5) & (6) \\
\midrule
 const & 32.607$^{***}$ & 32.607$^{***}$ & 32.607$^{***}$ & 29.703$^{***}$ & 29.703$^{***}$ & 29.703$^{***}$ \\
& (1.473) & (1.489) & (1.548) & (1.454) & (1.453) & (1.540) \\
 X\_dictator100 & 0.395$^{***}$ & & & 0.430$^{***}$ & & \\
& (0.067) & & & (0.069) & & \\
 X\_dictator300 & & 0.129$^{***}$ & & & 0.152$^{***}$ & \\
& & (0.023) & & & (0.024) & \\
 X\_dictator100x2 & & & & & & 0.243$^{***}$ \\
& & & & & & (0.071) \\
 X\_dictator100x0.5 & & & 0.205$^{***}$ & & & \\
& & & (0.063) & & & \\
\midrule
 Observations & 300 & 300 & 300 & 300 & 300 & 300 \\
 $R^2$ & 0.137 & 0.118 & 0.047 & 0.162 & 0.163 & 0.059 \\
\midrule
\multicolumn{7}{p{0.95\textwidth}}{\textit{Note:} Robust standard errors are reported in parentheses. $^{*}$p$<$0.1; $^{**}$p$<$0.05; $^{***}$p$<$0.01.} \\
\bottomrule
\end{tabular}
\label{tab:cross_give3} \end{table}

\begin{landscape}
    \begin{table}[!htbp] \centering
  \caption{Human Choices and Predictions About GenAI Choice in Related Problems}
  \begin{adjustbox}{scale=0.9}
\begin{tabular}{lcccccccccc}
\toprule
 & \multicolumn{5}{c}{P\_risk100} & \multicolumn{5}{c}{P\_risk200}  \\
\cmidrule(lr){2-6} \cmidrule(lr){7-11}
 & (1) & (2) & (3) & (4) & (5) & (6) & (7) & (8) & (9) & (10) \\
\midrule
 const & 36.483$^{***}$ & 36.483$^{***}$ & 36.483$^{***}$ & 36.483$^{***}$ & 36.483$^{***}$ & 96.843$^{***}$ & 96.843$^{***}$ & 96.843$^{***}$ & 96.843$^{***}$ & 96.843$^{***}$ \\
& (1.318) & (1.391) & (1.393) & (1.388) & (1.386) & (2.621) & (2.782) & (2.787) & (2.744) & (2.795) \\
 X\_risk100 & & & & & & 0.631$^{***}$ & & & & \\
& & & & & & (0.111) & & & & \\
 X\_risk200 & 0.157$^{***}$ & & & & & & & & & \\
& (0.032) & & & & & & & & & \\
 X\_dictator100 & & 0.153$^{**}$ & & & & & 0.264$^{**}$ & & & \\
& & (0.061) & & & & & (0.114) & & & \\
 X\_dictator300 & & & 0.050$^{**}$ & & & & & 0.085$^{**}$ & & \\
& & & (0.021) & & & & & (0.041) & & \\
 X\_dictator100x2 & & & & 0.153$^{**}$ & & & & & 0.378$^{***}$ & \\
& & & & (0.060) & & & & & (0.111) & \\
 X\_dictator100x0.5 & & & & & 0.154$^{***}$ & & & & & 0.170$^{}$ \\
& & & & & (0.052) & & & & & (0.107) \\
\midrule
 Observations & 300 & 300 & 300 & 300 & 300 & 300 & 300 & 300 & 300 & 300 \\
 $R^2$ & 0.126 & 0.026 & 0.023 & 0.030 & 0.033 & 0.129 & 0.019 & 0.016 & 0.046 & 0.010 \\
\midrule
\multicolumn{11}{p{0.95\textwidth}}{\textit{Note:} Robust standard errors are reported in parentheses. $^{*}$p$<$0.1; $^{**}$p$<$0.05; $^{***}$p$<$0.01.} \\
\bottomrule
\end{tabular}
\end{adjustbox}
\label{tab:cross_investment} \end{table}
\end{landscape}

\begin{landscape}
\subsection{Additional Tables for \cref{sec:human_on_human}}
\begin{table}[!htbp] \centering
  \caption{Predictions About Human Choice and Human Choices in the Same Problem}
  \begin{adjustbox}{scale=0.85}
\begin{tabular}{lccccccccc}
\toprule
 & \multicolumn{1}{c}{P\_risk100} & \multicolumn{1}{c}{P\_risk200} & \multicolumn{1}{c}{P\_discounting} & \multicolumn{1}{c}{P\_dictator100} & \multicolumn{1}{c}{P\_dictator300} & \multicolumn{1}{c}{P\_dictator100x2} & \multicolumn{1}{c}{P\_dictator100x0.5} & \multicolumn{1}{c}{P\_prisoner} & \multicolumn{1}{c}{P\_beauty}  \\
 & (1) & (2) & (3) & (4) & (5) & (6) & (7) & (8) & (9) \\
\midrule
 const & 33.163$^{***}$ & 84.017$^{***}$ & 290.627$^{***}$ & 23.927$^{***}$ & 67.913$^{***}$ & 24.523$^{***}$ & 22.760$^{***}$ & 56.297$^{***}$ & 48.537$^{***}$ \\
& (1.106) & (2.065) & (3.203) & (0.938) & (2.558) & (0.966) & (0.973) & (1.103) & (0.849) \\
 X & 0.334$^{***}$ & 0.356$^{***}$ & 0.568$^{***}$ & 0.551$^{***}$ & 0.579$^{***}$ & 0.474$^{***}$ & 0.590$^{***}$ & 0.223$^{***}$ & 0.442$^{***}$ \\
& (0.044) & (0.040) & (0.042) & (0.057) & (0.044) & (0.056) & (0.052) & (0.023) & (0.043) \\
\midrule
 Observations & 300 & 300 & 300 & 300 & 300 & 300 & 300 & 300 & 300 \\
 $R^2$ & 0.210 & 0.255 & 0.379 & 0.370 & 0.426 & 0.303 & 0.417 & 0.254 & 0.294 \\
\midrule
\multicolumn{10}{p{0.95\textwidth}}{\textit{Note:} $^{*}$p$<$0.1; $^{**}$p$<$0.05; $^{***}$p$<$0.01.} \\
\bottomrule
\end{tabular}
\end{adjustbox}
\label{tab:individual_regressions_sept} \end{table}

\end{landscape}

\begin{table}[!htbp] \centering
  \caption{Predictions About Human Choice and Human Choices in Related Problems}
\begin{tabular}{lcccccc}
\toprule
 & \multicolumn{3}{c}{P\_dictator100} & \multicolumn{3}{c}{P\_dictator300}  \\
\cmidrule(lr){2-4} \cmidrule(lr){5-7}
 & (1) & (2) & (3) & (4) & (5) & (6) \\
\midrule
 const & 23.927$^{***}$ & 23.927$^{***}$ & 23.927$^{***}$ & 67.913$^{***}$ & 67.913$^{***}$ & 67.913$^{***}$ \\
& (0.942) & (1.043) & (0.928) & (2.585) & (3.047) & (2.648) \\
 X\_dictator100 & & & & 1.667$^{***}$ & & \\
& & & & (0.161) & & \\
 X\_dictator300 & 0.187$^{***}$ & & & & & \\
& (0.016) & & & & & \\
 X\_dictator100x2 & & 0.411$^{***}$ & & & 1.081$^{***}$ & \\
& & (0.065) & & & (0.181) & \\
 X\_dictator100x0.5 & & & 0.524$^{***}$ & & & 1.502$^{***}$ \\
& & & (0.046) & & & (0.122) \\
\midrule
 Observations & 300 & 300 & 300 & 300 & 300 & 300 \\
 $R^2$ & 0.365 & 0.220 & 0.382 & 0.414 & 0.186 & 0.385 \\
\midrule
\multicolumn{7}{p{0.95\textwidth}}{\textit{Note:} $^{*}$p$<$0.1; $^{**}$p$<$0.05; $^{***}$p$<$0.01.} \\
\bottomrule
\end{tabular}
\label{tab:cross_give2_sept} \end{table}

\begin{table}[!htbp] \centering
  \caption{Predictions About Human Choice and Human Choices in Related Problems}
\begin{tabular}{lcccccc}
\toprule
 & \multicolumn{3}{c}{P\_dictator100x2} & \multicolumn{3}{c}{P\_dictator100x0.5}  \\
\cmidrule(lr){2-4} \cmidrule(lr){5-7}
 & (1) & (2) & (3) & (4) & (5) & (6) \\
\midrule
 const & 24.523$^{***}$ & 24.523$^{***}$ & 24.523$^{***}$ & 22.760$^{***}$ & 22.760$^{***}$ & 22.760$^{***}$ \\
& (1.012) & (1.022) & (1.024) & (1.077) & (1.084) & (1.203) \\
 X\_dictator100 & 0.432$^{***}$ & & & 0.521$^{***}$ & & \\
& (0.056) & & & (0.061) & & \\
 X\_dictator300 & & 0.143$^{***}$ & & & 0.176$^{***}$ & \\
& & (0.017) & & & (0.020) & \\
 X\_dictator100x2 & & & & & & 0.311$^{***}$ \\
& & & & & & (0.072) \\
 X\_dictator100x0.5 & & & 0.387$^{***}$ & & & \\
& & & (0.050) & & & \\
\midrule
 Observations & 300 & 300 & 300 & 300 & 300 & 300 \\
 $R^2$ & 0.236 & 0.220 & 0.217 & 0.284 & 0.276 & 0.108 \\
\midrule
\multicolumn{7}{p{0.95\textwidth}}{\textit{Note:} $^{*}$p$<$0.1; $^{**}$p$<$0.05; $^{***}$p$<$0.01.} \\
\bottomrule
\end{tabular}
\label{tab:cross_give3_sept} \end{table}

\begin{landscape}
    \begin{table}[!htbp] \centering
  \caption{Predictions About Human Choice and Human Choices in Related Problems}
\begin{adjustbox}{scale=0.85}
\begin{tabular}{lcccccccccc}
\toprule
 & \multicolumn{5}{c}{P\_risk100} & \multicolumn{5}{c}{P\_risk200}  \\
\cmidrule(lr){2-6} \cmidrule(lr){7-11}
 & (1) & (2) & (3) & (4) & (5) & (6) & (7) & (8) & (9) & (10) \\
\midrule
 const & 33.163$^{***}$ & 33.163$^{***}$ & 33.163$^{***}$ & 33.163$^{***}$ & 33.163$^{***}$ & 84.017$^{***}$ & 84.017$^{***}$ & 84.017$^{***}$ & 84.017$^{***}$ & 84.017$^{***}$ \\
& (1.189) & (1.229) & (1.222) & (1.241) & (1.223) & (2.262) & (2.374) & (2.383) & (2.364) & (2.385) \\
 X\_risk100 & & & & & & 0.457$^{***}$ & & & & \\
& & & & & & (0.089) & & & & \\
 X\_risk200 & 0.108$^{***}$ & & & & & & & & & \\
& (0.023) & & & & & & & & & \\
 X\_dictator100 & & 0.152$^{***}$ & & & & & 0.226$^{**}$ & & & \\
& & (0.059) & & & & & (0.105) & & & \\
 X\_dictator300 & & & 0.062$^{***}$ & & & & & 0.056$^{}$ & & \\
& & & (0.020) & & & & & (0.035) & & \\
 X\_dictator100x2 & & & & 0.073$^{}$ & & & & & 0.272$^{***}$ & \\
& & & & (0.058) & & & & & (0.104) & \\
 X\_dictator100x0.5 & & & & & 0.167$^{***}$ & & & & & 0.136$^{}$ \\
& & & & & (0.052) & & & & & (0.097) \\
\midrule
 Observations & 300 & 300 & 300 & 300 & 300 & 300 & 300 & 300 & 300 & 300 \\
 $R^2$ & 0.087 & 0.025 & 0.036 & 0.006 & 0.035 & 0.106 & 0.015 & 0.008 & 0.023 & 0.006 \\
\midrule
\multicolumn{11}{p{0.95\textwidth}}{\textit{Note:} $^{*}$p$<$0.1; $^{**}$p$<$0.05; $^{***}$p$<$0.01.} \\
\bottomrule
\end{tabular}
\end{adjustbox}
\label{tab:cross_investment_sept} \end{table}

\end{landscape}

\begin{table}[!htbp] \centering
  \caption{Predictions About Human Choice and Human Choices in the Same Problem: Heterogeneity by Response Time (Pooled)}
\begin{tabular}{lcc}
\toprule
& \multicolumn{2}{c}{Dependent variable: Prediction} \\
\cmidrule(lr){2-3}
 & \multicolumn{1}{c}{Problem Median} & \multicolumn{1}{c}{Personal Median}  \\
 & (1) & (2) \\
\midrule
 X$\times$T & 0.006$^{}$ & 0.005$^{}$ \\
& (0.025) & (0.027) \\
 Problem FE & Yes & Yes \\
 X$\times$Problem FE & Yes & Yes \\
 T$\times$Problem FE & Yes & Yes \\
\midrule
 Observations & 2700 & 2700 \\
\midrule
\multicolumn{3}{p{0.95\textwidth}}{\textit{Note:} Standard errors are clustered at the problem level. $^{*}$p$<$0.1; $^{**}$p$<$0.05; $^{***}$p$<$0.01.} \\
\bottomrule
\end{tabular}
\label{tab:time_sept} \end{table}

\begin{table}[!htbp] \centering
  \caption{Gender}
\begin{tabular}{lc}
\toprule
& \multicolumn{1}{c}{Dependent variable: Prediction} \\
\cmidrule(lr){2-2}
 & (1) \\
\midrule
 X$\times$Female & 0.003$^{}$ \\
& (0.032) \\
 Problem FE & Yes \\
 X$\times$Problem FE & Yes \\
 Female$\times$Problem FE & Yes \\
\midrule
 Observations & 2682 \\
\midrule
\multicolumn{2}{p{0.95\textwidth}}{\textit{Note:} Standard errors are clustered at the problem level. $^{*}$p$<$0.1; $^{**}$p$<$0.05; $^{***}$p$<$0.01.} \\
\bottomrule
\end{tabular}
\label{tab:gender_sept} \end{table}

\begin{landscape}
\subsection{Additional Tables for \cref{subsec:interact_GenAI_exp}}\label{Additional: experience}

    \begin{table}[!htbp] \centering
  \caption{Human Choices and Predictions About GenAI Choice in the Same Problem: Heterogeneity by GenAI Usage}
\begin{adjustbox}{scale=0.85}
\begin{tabular}{lccccccccc}
\toprule
 & \multicolumn{1}{c}{P\_risk100} & \multicolumn{1}{c}{P\_risk200} & \multicolumn{1}{c}{P\_discounting} & \multicolumn{1}{c}{P\_dictator100} & \multicolumn{1}{c}{P\_dictator300} & \multicolumn{1}{c}{P\_dictator100x2} & \multicolumn{1}{c}{P\_dictator100x0.5} & \multicolumn{1}{c}{P\_prisoner} & \multicolumn{1}{c}{P\_beauty}  \\
 & (1) & (2) & (3) & (4) & (5) & (6) & (7) & (8) & (9) \\
\midrule
 const & 35.413$^{***}$ & 95.094$^{***}$ & 286.696$^{***}$ & 30.487$^{***}$ & 88.168$^{***}$ & 32.196$^{***}$ & 29.992$^{***}$ & 51.240$^{***}$ & 47.886$^{***}$ \\
& (1.906) & (3.034) & (4.796) & (2.011) & (5.385) & (1.893) & (1.996) & (2.070) & (1.228) \\
 X & 0.239$^{***}$ & 0.380$^{***}$ & 0.520$^{***}$ & 0.437$^{***}$ & 0.481$^{***}$ & 0.612$^{***}$ & 0.497$^{***}$ & 0.146$^{***}$ & 0.515$^{***}$ \\
& (0.090) & (0.069) & (0.079) & (0.102) & (0.093) & (0.087) & (0.083) & (0.041) & (0.065) \\
 Heavy User & 2.566$^{}$ & 3.798$^{}$ & -9.953$^{}$ & 2.476$^{}$ & 4.614$^{}$ & 0.940$^{}$ & -0.532$^{}$ & -0.123$^{}$ & 1.500$^{}$ \\
& (2.526) & (4.858) & (7.040) & (2.784) & (7.744) & (2.749) & (2.894) & (2.879) & (1.949) \\
 X$\times$Heavy User & 0.268$^{**}$ & 0.126$^{}$ & -0.125$^{}$ & -0.180$^{}$ & -0.089$^{}$ & -0.233$^{*}$ & -0.234$^{*}$ & 0.006$^{}$ & -0.223$^{**}$ \\
& (0.111) & (0.103) & (0.109) & (0.140) & (0.130) & (0.120) & (0.121) & (0.057) & (0.105) \\
\midrule
 Observations & 300 & 300 & 300 & 300 & 300 & 300 & 300 & 300 & 300 \\
 $R^2$ & 0.201 & 0.258 & 0.250 & 0.130 & 0.187 & 0.259 & 0.177 & 0.081 & 0.260 \\
\midrule
\multicolumn{10}{p{0.95\textwidth}}{\textit{Note:} Robust standard errors are reported in parentheses. $^{*}$p$<$0.1; $^{**}$p$<$0.05; $^{***}$p$<$0.01.} \\
\bottomrule
\end{tabular}
\end{adjustbox}
\label{tab:group_heavy_user} \end{table}

\end{landscape}

\begin{landscape}
     \begin{table}[!htbp] \centering
  \caption{Human Choices and Predictions About GenAI Choice in the Same Problem: Heterogeneity by LLM Usage}
\begin{adjustbox}{scale=0.8}
\begin{tabular}{lccccccccc}
\toprule
 & \multicolumn{1}{c}{P\_risk100} & \multicolumn{1}{c}{P\_risk200} & \multicolumn{1}{c}{P\_discounting} & \multicolumn{1}{c}{P\_dictator100} & \multicolumn{1}{c}{P\_dictator300} & \multicolumn{1}{c}{P\_dictator100x2} & \multicolumn{1}{c}{P\_dictator100x0.5} & \multicolumn{1}{c}{P\_prisoner} & \multicolumn{1}{c}{P\_beauty}  \\
 & (1) & (2) & (3) & (4) & (5) & (6) & (7) & (8) & (9) \\
\midrule
 const & 29.781$^{***}$ & 87.063$^{***}$ & 294.297$^{***}$ & 28.435$^{***}$ & 91.681$^{***}$ & 29.686$^{***}$ & 30.065$^{***}$ & 44.810$^{***}$ & 50.637$^{***}$ \\
& (3.860) & (8.836) & (9.276) & (4.842) & (10.919) & (5.264) & (4.581) & (4.340) & (3.580) \\
 X & 0.053$^{}$ & 0.290$^{}$ & 0.357$^{***}$ & 0.474$^{}$ & 0.840$^{***}$ & 0.377$^{*}$ & 0.315$^{}$ & 0.080$^{}$ & 0.267$^{}$ \\
& (0.184) & (0.180) & (0.133) & (0.305) & (0.137) & (0.203) & (0.212) & (0.086) & (0.194) \\
 Text-Based LLM User & 7.374$^{*}$ & 10.786$^{}$ & -13.966$^{}$ & 3.736$^{}$ & -0.128$^{}$ & 3.223$^{}$ & -0.436$^{}$ & 7.303$^{}$ & -2.341$^{}$ \\
& (4.085) & (9.194) & (9.991) & (5.057) & (11.680) & (5.450) & (4.830) & (4.596) & (3.717) \\
 X$\times$Text-Based LLM User & 0.337$^{*}$ & 0.165$^{}$ & 0.124$^{}$ & -0.142$^{}$ & -0.440$^{***}$ & 0.126$^{}$ & 0.075$^{}$ & 0.081$^{}$ & 0.154$^{}$ \\
& (0.194) & (0.188) & (0.145) & (0.314) & (0.153) & (0.212) & (0.222) & (0.091) & (0.202) \\
\midrule
 Observations & 300 & 300 & 300 & 300 & 300 & 300 & 300 & 300 & 300 \\
 $R^2$ & 0.192 & 0.258 & 0.248 & 0.124 & 0.197 & 0.248 & 0.162 & 0.093 & 0.246 \\
\midrule
\multicolumn{10}{p{0.95\textwidth}}{\textit{Note:} Robust standard errors are reported in parentheses. $^{*}$p$<$0.1; $^{**}$p$<$0.05; $^{***}$p$<$0.01.} \\
\bottomrule
\end{tabular}
\end{adjustbox}
\label{tab:group_llm_user} \end{table}
    \begin{table}[!htbp] \centering
  \caption{Human Choices and Predictions About GenAI Choice in the Same Problem: Heterogeneity by Paid Usage}
\begin{adjustbox}{scale=0.85}
\begin{tabular}{lccccccccc}
\toprule
 & \multicolumn{1}{c}{P\_risk100} & \multicolumn{1}{c}{P\_risk200} & \multicolumn{1}{c}{P\_discounting} & \multicolumn{1}{c}{P\_dictator100} & \multicolumn{1}{c}{P\_dictator300} & \multicolumn{1}{c}{P\_dictator100x2} & \multicolumn{1}{c}{P\_dictator100x0.5} & \multicolumn{1}{c}{P\_prisoner} & \multicolumn{1}{c}{P\_beauty}  \\
 & (1) & (2) & (3) & (4) & (5) & (6) & (7) & (8) & (9) \\
\midrule
 const & 36.204$^{***}$ & 95.374$^{***}$ & 282.411$^{***}$ & 32.045$^{***}$ & 91.891$^{***}$ & 32.831$^{***}$ & 30.741$^{***}$ & 51.164$^{***}$ & 48.740$^{***}$ \\
& (1.371) & (2.513) & (3.675) & (1.464) & (4.071) & (1.449) & (1.535) & (1.514) & (1.025) \\
 X & 0.364$^{***}$ & 0.453$^{***}$ & 0.430$^{***}$ & 0.375$^{***}$ & 0.462$^{***}$ & 0.510$^{***}$ & 0.411$^{***}$ & 0.139$^{***}$ & 0.361$^{***}$ \\
& (0.064) & (0.055) & (0.058) & (0.072) & (0.065) & (0.061) & (0.064) & (0.030) & (0.057) \\
 Paid User & 2.945$^{}$ & 15.085$^{}$ & 0.738$^{}$ & -3.611$^{}$ & -15.933$^{}$ & -2.936$^{}$ & -9.792$^{**}$ & 1.024$^{}$ & -1.524$^{}$ \\
& (3.691) & (9.219) & (12.207) & (4.710) & (12.463) & (4.897) & (4.306) & (5.283) & (3.124) \\
 X$\times$Paid User & 0.050$^{}$ & -0.101$^{}$ & 0.227$^{}$ & -0.276$^{}$ & -0.290$^{}$ & -0.218$^{}$ & -0.262$^{}$ & 0.103$^{}$ & 0.403$^{***}$ \\
& (0.147) & (0.170) & (0.160) & (0.236) & (0.227) & (0.268) & (0.203) & (0.097) & (0.140) \\
\midrule
 Observations & 300 & 300 & 300 & 300 & 300 & 300 & 300 & 300 & 300 \\
 $R^2$ & 0.176 & 0.261 & 0.247 & 0.129 & 0.195 & 0.249 & 0.181 & 0.084 & 0.262 \\
\midrule
\multicolumn{10}{p{0.95\textwidth}}{\textit{Note:} Robust standard errors are reported in parentheses. $^{*}$p$<$0.1; $^{**}$p$<$0.05; $^{***}$p$<$0.01.} \\
\bottomrule
\end{tabular}
\end{adjustbox}
\label{tab:group_paid_user} \end{table}
\end{landscape}

\begin{landscape}
    \begin{table}[!htbp] \centering
  \caption{Human Choices and Predictions About GenAI Choice in the Same Problem: Heterogeneity by Attitude (``AI Similar")  }
\begin{adjustbox}{scale=0.8}
\begin{tabular}{lccccccccc}
\toprule
 & \multicolumn{1}{c}{P\_risk100} & \multicolumn{1}{c}{P\_risk200} & \multicolumn{1}{c}{P\_discounting} & \multicolumn{1}{c}{P\_dictator100} & \multicolumn{1}{c}{P\_dictator300} & \multicolumn{1}{c}{P\_dictator100x2} & \multicolumn{1}{c}{P\_dictator100x0.5} & \multicolumn{1}{c}{P\_prisoner} & \multicolumn{1}{c}{P\_beauty}  \\
 & (1) & (2) & (3) & (4) & (5) & (6) & (7) & (8) & (9) \\
\midrule
 const & 36.425$^{***}$ & 94.704$^{***}$ & 284.314$^{***}$ & 30.212$^{***}$ & 83.343$^{***}$ & 32.017$^{***}$ & 28.386$^{***}$ & 50.268$^{***}$ & 49.300$^{***}$ \\
& (1.676) & (3.041) & (4.417) & (1.727) & (4.768) & (1.752) & (1.892) & (1.763) & (1.183) \\
 X & 0.327$^{***}$ & 0.487$^{***}$ & 0.416$^{***}$ & 0.433$^{***}$ & 0.416$^{***}$ & 0.499$^{***}$ & 0.351$^{***}$ & 0.106$^{***}$ & 0.447$^{***}$ \\
& (0.078) & (0.063) & (0.067) & (0.078) & (0.075) & (0.073) & (0.081) & (0.035) & (0.067) \\
 Agree AI Similar & 0.032$^{}$ & 5.732$^{}$ & -4.830$^{}$ & 4.441$^{}$ & 20.571$^{**}$ & 1.733$^{}$ & 3.730$^{}$ & 3.047$^{}$ & -2.084$^{}$ \\
& (2.550) & (5.032) & (7.213) & (2.871) & (8.098) & (2.837) & (2.918) & (3.041) & (2.093) \\
 X$\times$Agree AI Similar & 0.122$^{}$ & -0.133$^{}$ & 0.123$^{}$ & -0.249$^{*}$ & 0.073$^{}$ & -0.018$^{}$ & 0.088$^{}$ & 0.125$^{**}$ & -0.116$^{}$ \\
& (0.114) & (0.112) & (0.113) & (0.148) & (0.145) & (0.131) & (0.121) & (0.059) & (0.110) \\
\midrule
 Observations & 300 & 300 & 300 & 300 & 300 & 300 & 300 & 300 & 300 \\
 $R^2$ & 0.179 & 0.259 & 0.246 & 0.141 & 0.203 & 0.246 & 0.168 & 0.097 & 0.248 \\
\midrule
\multicolumn{10}{p{0.95\textwidth}}{\textit{Note:} Robust standard errors are reported in parentheses. $^{*}$p$<$0.1; $^{**}$p$<$0.05; $^{***}$p$<$0.01.} \\
\bottomrule
\end{tabular}
\end{adjustbox}
\label{tab:group_ai_similar} \end{table}
    \begin{table}[!htbp]  \centering
  \caption{Human Choices and Predictions About GenAI Choice in the Same Problem: Heterogeneity by Attitude (``AI Better")}
\begin{adjustbox}{scale=0.8}
\begin{tabular}{lccccccccc}
\toprule
 & \multicolumn{1}{c}{P\_risk100} & \multicolumn{1}{c}{P\_risk200} & \multicolumn{1}{c}{P\_discounting} & \multicolumn{1}{c}{P\_dictator100} & \multicolumn{1}{c}{P\_dictator300} & \multicolumn{1}{c}{P\_dictator100x2} & \multicolumn{1}{c}{P\_dictator100x0.5} & \multicolumn{1}{c}{P\_prisoner} & \multicolumn{1}{c}{P\_beauty}  \\
 & (1) & (2) & (3) & (4) & (5) & (6) & (7) & (8) & (9) \\
\midrule
 const & 36.275$^{***}$ & 98.000$^{***}$ & 283.298$^{***}$ & 31.460$^{***}$ & 87.128$^{***}$ & 32.535$^{***}$ & 29.717$^{***}$ & 51.822$^{***}$ & 48.072$^{***}$ \\
& (1.554) & (2.892) & (4.135) & (1.687) & (4.684) & (1.675) & (1.773) & (1.627) & (1.124) \\
 X & 0.348$^{***}$ & 0.449$^{***}$ & 0.475$^{***}$ & 0.340$^{***}$ & 0.391$^{***}$ & 0.480$^{***}$ & 0.392$^{***}$ & 0.138$^{***}$ & 0.410$^{***}$ \\
& (0.069) & (0.063) & (0.067) & (0.090) & (0.081) & (0.077) & (0.073) & (0.032) & (0.064) \\
 Agree AI Better & 0.671$^{}$ & -4.325$^{}$ & -4.468$^{}$ & 0.829$^{}$ & 11.510$^{}$ & 0.264$^{}$ & -0.042$^{}$ & -2.280$^{}$ & 1.922$^{}$ \\
& (2.685) & (5.342) & (7.680) & (2.978) & (8.156) & (2.896) & (3.033) & (3.465) & (2.284) \\
 X$\times$Agree AI Better & 0.075$^{}$ & -0.017$^{}$ & -0.058$^{}$ & 0.022$^{}$ & 0.138$^{}$ & 0.049$^{}$ & -0.039$^{}$ & 0.034$^{}$ & -0.040$^{}$ \\
& (0.132) & (0.113) & (0.115) & (0.134) & (0.120) & (0.118) & (0.138) & (0.067) & (0.121) \\
\midrule
 Observations & 300 & 300 & 300 & 300 & 300 & 300 & 300 & 300 & 300 \\
 $R^2$ & 0.177 & 0.253 & 0.242 & 0.120 & 0.193 & 0.246 & 0.162 & 0.083 & 0.243 \\
\midrule
\multicolumn{10}{p{0.95\textwidth}}{\textit{Note:} Robust standard errors are reported in parentheses. $^{*}$p$<$0.1; $^{**}$p$<$0.05; $^{***}$p$<$0.01.} \\
\bottomrule
\end{tabular}
\end{adjustbox}
\label{tab:group_ai_better}
\end{table}
\end{landscape}

\begin{table}[!htbp] \centering
  \caption{Individual Prediction Accuracy and GenAI Experience, GPT-4o}
\begin{tabular}{lccc}
\toprule
& \multicolumn{3}{c}{Dependent variable: MAE} \\
\cmidrule(lr){2-4}
 & (1) & (2) & (3) \\
\midrule
 const & 33.92$^{***}$ & 35.04$^{***}$ & 33.80$^{***}$ \\
& (1.07) & (1.72) & (0.93) \\
 Heavy User & -0.13$^{}$ & & \\
& (1.33) & & \\
 GPT User & & -1.42$^{}$ & \\
& & (1.85) & \\
 Paid User & & & 2.90$^{}$ \\
& & & (2.34) \\
 Agree AI Similar & -2.80$^{**}$ & -2.74$^{**}$ & -3.10$^{**}$ \\
& (1.31) & (1.33) & (1.32) \\
 Agree AI Better & -1.29$^{}$ & -1.21$^{}$ & -1.53$^{}$ \\
& (1.47) & (1.44) & (1.42) \\
 Copier & -5.63$^{***}$ & -5.54$^{***}$ & -6.18$^{***}$ \\
& (1.98) & (1.95) & (1.94) \\
\midrule
 Observations & 300 & 300 & 300 \\
 $R^2$ & 0.05 & 0.05 & 0.05 \\
\midrule
\multicolumn{4}{p{0.95\textwidth}}{\textit{Note:} Robust standard errors are reported in parentheses. $^{*}$p$<$0.1; $^{**}$p$<$0.05; $^{***}$p$<$0.01.} \\
\bottomrule
\end{tabular}
\label{tab:mae_gpt-4o} \end{table}

\begin{table}[!htbp] \centering
  \caption{Individual Prediction Accuracy and GenAI Experience, GPT-4o-mini}
\begin{tabular}{lccc}
\toprule
& \multicolumn{3}{c}{Dependent variable: MAE} \\
\cmidrule(lr){2-4}
 & (1) & (2) & (3) \\
\midrule
 const & 27.06$^{***}$ & 28.61$^{***}$ & 27.36$^{***}$ \\
& (0.98) & (1.53) & (0.82) \\
 Heavy User & 0.94$^{}$ & & \\
& (1.03) & & \\
 GPT User & & -1.49$^{}$ & \\
& & (1.59) & \\
 Paid User & & & 1.07$^{}$ \\
& & & (1.25) \\
 Agree AI Similar & -0.96$^{}$ & -0.76$^{}$ & -0.95$^{}$ \\
& (0.99) & (1.00) & (1.02) \\
 Agree AI Better & -0.94$^{}$ & -0.58$^{}$ & -0.78$^{}$ \\
& (1.12) & (1.13) & (1.14) \\
 Copier & 0.80$^{}$ & 1.24$^{}$ & 0.91$^{}$ \\
& (1.36) & (1.38) & (1.30) \\
\midrule
 Observations & 300 & 300 & 300 \\
 $R^2$ & 0.01 & 0.01 & 0.01 \\
\midrule
\multicolumn{4}{p{0.95\textwidth}}{\textit{Note:} Robust standard errors are reported in parentheses. $^{*}$p$<$0.1; $^{**}$p$<$0.05; $^{***}$p$<$0.01.} \\
\bottomrule
\end{tabular}
\label{tab:mae_gpt-4o-mini} \end{table}

\begin{table}[!htbp] \centering
  \caption{Individual Prediction Accuracy and GenAI Experience, Gemini-1.5-Pro}
\begin{tabular}{lccc}
\toprule
& \multicolumn{3}{c}{Dependent variable: MAE} \\
\cmidrule(lr){2-4}
 & (1) & (2) & (3) \\
\midrule
 const & 50.65$^{***}$ & 51.25$^{***}$ & 50.66$^{***}$ \\
& (1.22) & (1.13) & (1.08) \\
 Heavy User & 0.35$^{}$ & & \\
& (1.62) & & \\
 Gemini User & & -1.77$^{}$ & \\
& & (1.65) & \\
 Paid User & & & 4.57$^{*}$ \\
& & & (2.77) \\
 Agree AI Similar & -3.11$^{**}$ & -2.77$^{*}$ & -3.52$^{**}$ \\
& (1.57) & (1.61) & (1.56) \\
 Agree AI Better & -0.96$^{}$ & -0.64$^{}$ & -1.19$^{}$ \\
& (1.72) & (1.71) & (1.67) \\
 Copier & -8.14$^{***}$ & -7.93$^{***}$ & -8.82$^{***}$ \\
& (2.15) & (2.13) & (2.15) \\
\midrule
 Observations & 300 & 300 & 300 \\
 $R^2$ & 0.05 & 0.06 & 0.06 \\
\midrule
\multicolumn{4}{p{0.95\textwidth}}{\textit{Note:} Robust standard errors are reported in parentheses. $^{*}$p$<$0.1; $^{**}$p$<$0.05; $^{***}$p$<$0.01.} \\
\bottomrule
\end{tabular}
\label{tab:mae_gemini-1.5-pro} \end{table}

\begin{table}[!htbp] \centering
  \caption{Individual Prediction Accuracy and GenAI Experience, Gemini-1.5-Flash}
\begin{tabular}{lccc}
\toprule
& \multicolumn{3}{c}{Dependent variable: MAE} \\
\cmidrule(lr){2-4}
 & (1) & (2) & (3) \\
\midrule
 const & 31.44$^{***}$ & 31.48$^{***}$ & 31.47$^{***}$ \\
& (0.79) & (0.69) & (0.67) \\
 Heavy User & 0.14$^{}$ & & \\
& (0.87) & & \\
 Gemini User & & 0.01$^{}$ & \\
& & (0.96) & \\
 Paid User & & & 0.83$^{}$ \\
& & & (1.19) \\
 Agree AI Similar & -0.97$^{}$ & -0.96$^{}$ & -1.04$^{}$ \\
& (0.83) & (0.86) & (0.86) \\
 Agree AI Better & 0.06$^{}$ & 0.10$^{}$ & 0.04$^{}$ \\
& (0.94) & (0.94) & (0.93) \\
 Copier & -3.34$^{**}$ & -3.30$^{**}$ & -3.44$^{**}$ \\
& (1.45) & (1.45) & (1.47) \\
\midrule
 Observations & 300 & 300 & 300 \\
 $R^2$ & 0.02 & 0.02 & 0.02 \\
\midrule
\multicolumn{4}{p{0.95\textwidth}}{\textit{Note:} Robust standard errors are reported in parentheses. $^{*}$p$<$0.1; $^{**}$p$<$0.05; $^{***}$p$<$0.01.} \\
\bottomrule
\end{tabular}
\label{tab:mae_gemini-1.5-flash} \end{table}

\clearpage
\subsection{Additional Tables for \cref{subsec:response time}}
\begin{table}[!htbp] \centering
  \caption{Human Choices and Predictions About GenAI Choice in the Same Problem: Heterogeneity by Response Time (Pooled)}
\begin{tabular}{lcc}
\toprule
& \multicolumn{2}{c}{Dependent variable: Prediction} \\
\cmidrule(lr){2-3}
 & \multicolumn{1}{c}{Problem Median} & \multicolumn{1}{c}{Personal Median}  \\
 & (1) & (2) \\
\midrule
 X$\times$T & -0.055$^{**}$ & -0.048$^{}$ \\
& (0.024) & (0.036) \\
 Problem FE & Yes & Yes \\
 X$\times$Problem FE & Yes & Yes \\
 T$\times$Problem FE & Yes & Yes \\
\midrule
 Observations & 2700 & 2700 \\
\midrule
\multicolumn{3}{p{0.95\textwidth}}{\textit{Note:} Standard errors are clustered at the problem level. $^{*}$p$<$0.1; $^{**}$p$<$0.05; $^{***}$p$<$0.01.} \\
\bottomrule
\end{tabular}
\label{tab:time} \end{table}
\subsection{Additional Tables for \cref{subsec: gender}}

\begin{table}[!htbp] \centering
  \caption{Human Choices and Predictions About GenAI Choice in the Same Problem: Heterogeneity by Gender (Pooled)}
\begin{tabular}{lc}
\toprule
& \multicolumn{1}{c}{Dependent variable: Prediction} \\
\cmidrule(lr){2-2}
 & (1) \\
\midrule
 X$\times$Female & 0.005$^{}$ \\
& (0.018) \\
 Problem FE & Yes \\
 X$\times$Problem FE & Yes \\
 Female$\times$Problem FE & Yes \\
\midrule
 Observations & 2646 \\
\midrule
\multicolumn{2}{p{0.95\textwidth}}{\textit{Note:} Standard errors are clustered at the problem level. $^{*}$p$<$0.1; $^{**}$p$<$0.05; $^{***}$p$<$0.01.} \\
\bottomrule
\end{tabular}
\label{tab:gender} \end{table}

\clearpage
\section{Prompts} \label{sec:prompts}

The following paragraph is used as developer instructions in GPT and as system instructions in Gemini.
\begin{quote}
You are a powerful decision-making agent and a helpful assistant that strictly follows the user’s instructions. The user is busy and requires you to provide an answer in exactly the requested format. The user may be given tokens depending on the answer you provide; each token is worth 0.001 US dollars. Here is the question that the user is facing:
\end{quote}

The prompts given to human subjects and GenAI models are nearly identical. Sentences included only in the GenAI prompts will be underlined. For human subjects, we emphasized the important parts of the problem using bold text, as shown below.

Prompt for \textbf{risk100}
\begin{quote}
    You have \textbf{100} tokens. Please choose how many tokens out of the \textbf{100} to invest. The tokens you invest will be taken away, and you get to keep all the tokens that you choose not to invest. With \textbf{35}\% probability, the investment will be successful, and you will receive \textbf{3} tokens for every token that you invested. With \textbf{65}\% probability, the investment will be unsuccessful, and you will not receive anything for the tokens that you invested. How many tokens do you choose to invest? \uline{Your answer must contain only a number, nothing else. Answer:}
\end{quote}

Prompt for \textbf{risk200}
\begin{quote}
    You have \textbf{200} tokens. Please choose how many tokens out of the \textbf{200} to invest. The tokens you invest will be taken away, and you get to keep all the tokens that you choose not to invest. With \textbf{50}\% probability, the investment will be successful, and you will receive \textbf{2.5} tokens for every token that you invested. With \textbf{50}\% probability, the investment will be unsuccessful, and you will not receive anything for the tokens that you invested. How many tokens do you choose to invest? \uline{Your answer must contain only a number, nothing else. Answer:}
\end{quote}

Prompt for \textbf{discounting}
\begin{quote}
    After this study ends and you receive your base payment and bonus payment, you will also receive an additional bonus payment in either 30 days or 60 days. One option is to receive 150 tokens (to be converted into dollars) in 30 days. Another option is to receive a larger number of tokens (again, to be converted into dollars) in 60 days. \textbf{How many tokens do we need to give you in 60 days to make that option as good for you as getting 150 tokens in 30 days?} Enter a number between 150 and 400.

    It is in your interest to answer accurately. After you enter your answer below (for example, let’s say you answer that N tokens in 60 days is as good as 150 tokens in 30 days), the computer will randomly draw a number X between 150 and 400, and this will be the number of tokens associated with the 60-days option. The computer will then choose between the option of “150 tokens in 30 days” and the option of “X tokens in 60 days”, based on your answer. If X is larger than N, then you will receive X tokens in 60 days. If X is smaller than N, then you will receive 150 tokens in 30 days. So, you will always get the option that you like better by accurately reporting how many tokens received in 60 days is equivalent (for you) compared to 150 tokens received in 30 days.

    Please enter below \textbf{how many tokens we need to give you in 60 days to make that option as good for you as getting 150 tokens in 30 days}. \uline{Your answer must contain only a number, nothing else. Answer:}
\end{quote}

Prompt for \textbf{dictator100}
\begin{quote}
    You have \textbf{100} tokens. The computer has paired you with another randomly selected Prolific participant from this study. You must choose how many tokens out of the \textbf{100} to give away. The tokens that you do not give away are yours to keep. For each token that you give away, \textbf{the other participant will receive one token}. These received tokens will be converted into dollars and paid to the other participant as an extra bonus payment. How many tokens will you give away? \uline{Your answer must contain only a number, nothing else. Answer:}
\end{quote}

Prompt for \textbf{dictator300}
\begin{quote}
    You have \textbf{300} tokens. The computer has paired you with another randomly selected Prolific participant from this study. You must choose how many tokens out of the \textbf{300} to give away. The tokens that you do not give away are yours to keep. For each token that you give away, \textbf{the other participant will receive one token}. These received tokens will be converted into dollars and paid to the other participant as an extra bonus payment. How many tokens will you give away? \uline{Your answer must contain only a number, nothing else. Answer:}
\end{quote}

Prompt for \textbf{dictator100x2}
\begin{quote}
    You have \textbf{100} tokens. The computer has paired you with another randomly selected Prolific participant from this study. You must choose how many tokens out of the \textbf{100} to give away. The tokens that you do not give away are yours to keep. For each token that you give away, \textbf{the other participant will receive two tokens}. These received tokens will be converted into dollars and paid to the other participant as an extra bonus payment. How many tokens will you give away? \uline{Your answer must contain only a number, nothing else. Answer:}
\end{quote}

Prompt for \textbf{dictator100x0.5}
\begin{quote}
    You have \textbf{100} tokens. The computer has paired you with another randomly selected Prolific participant from this study. You must choose how many tokens out of the \textbf{100} to give away. The tokens that you do not give away are yours to keep. For each token that you give away, \textbf{the other participant will receive half of a token}. These received tokens will be converted into dollars and paid to the other participant as an extra bonus payment. How many tokens will you give away? \uline{Your answer must contain only a number, nothing else. Answer:}
\end{quote}

For \textbf{prisoner}, the prompts differ from the human subjects instructions. Human subjects are presented with a table of payoffs, while GenAI is provided with a description of the payoffs for different outcomes.

Human version:
\begin{quote}
    The computer has paired you with another randomly selected Prolific participant from this study to play a game. Each of you will choose A or B.

    \begin{table}[h!]
    \centering
    \begin{tabular}{|>{\centering\arraybackslash}m{5cm}|>{\centering\arraybackslash}m{4cm}|>{\centering\arraybackslash}m{4cm}|}
    \hline
    If... & \multicolumn{2}{c|}{\begin{tabular}[c]{@{}c@{}} Then: \\ You get \qquad \qquad \qquad They get \end{tabular}} \\ \hline
    You choose A and they choose A & 80 tokens & 80 tokens \\ \hline
    You choose A and they choose B & 60 tokens & 90 tokens \\ \hline
    You choose B and they choose A & 90 tokens & 60 tokens \\ \hline
    You choose B and they choose B & 70 tokens & 70 tokens \\ \hline
    \end{tabular}
    \end{table}

    Do you choose A or B?
\end{quote}

GenAI version:
\begin{quote}
    The computer has paired you with another randomly selected Prolific participant from this study to play a game. Each of you will choose A or B. If you choose A and the other person chooses A, then you will get 80 tokens and the other person will get 80 tokens. If you choose A and the other person chooses B, then you will get 60 tokens and the other person will get 90 tokens. If you choose B and the other person chooses A, then you will get 90 tokens and the other person will get 60 tokens. If you choose B and the other person chooses B, then you will get 70 tokens and the other person will get 70 tokens. Do you choose A or B? \uline{Your answer must be either A or B and must contain nothing else. Answer:}
\end{quote}

Prompt for \textbf{beauty}
\begin{quote}
    You will play a guessing game with all other Prolific participants from this study. Everyone will enter a whole number between 0 and 100. The person whose number is the closest to \textbf{two-thirds of the average of the numbers} entered by all participants will win 5000 tokens. (If there is a tie for the closest number, then a winner will be randomly chosen among those who entered the closest number.) Enter your number below. \uline{Your answer must contain only a number, nothing else. Answer:}
\end{quote}

Prompt for \textbf{dictator200}
\begin{quote}
    You have 200 tokens. The computer has paired you with another randomly selected Prolific participant from this study. You must choose how many tokens out of the 200 to give away. The tokens that you do not give away are yours to keep. For each token that you give away, the other participant will receive one token. These received tokens will be converted into dollars and paid to the other participant as an extra bonus payment. How many tokens will you give away? Your answer must contain only a number, nothing else. Answer:
\end{quote}

\newpage
\section{Screenshots from User Interface}

\begin{figure}[h]
    \centering
    \includegraphics[width=0.9\linewidth]{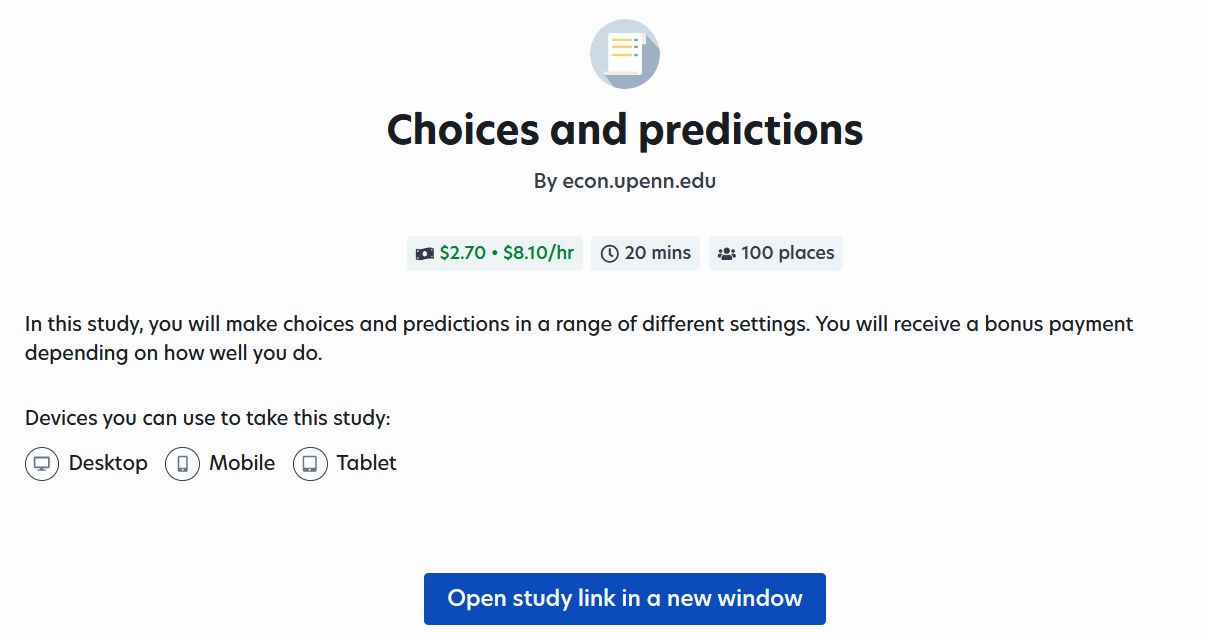}
    \caption{Launch Page}
    \label{fig:landing}
\end{figure}

\begin{figure}[h]
    \centering
    \includegraphics[width=0.9\linewidth]{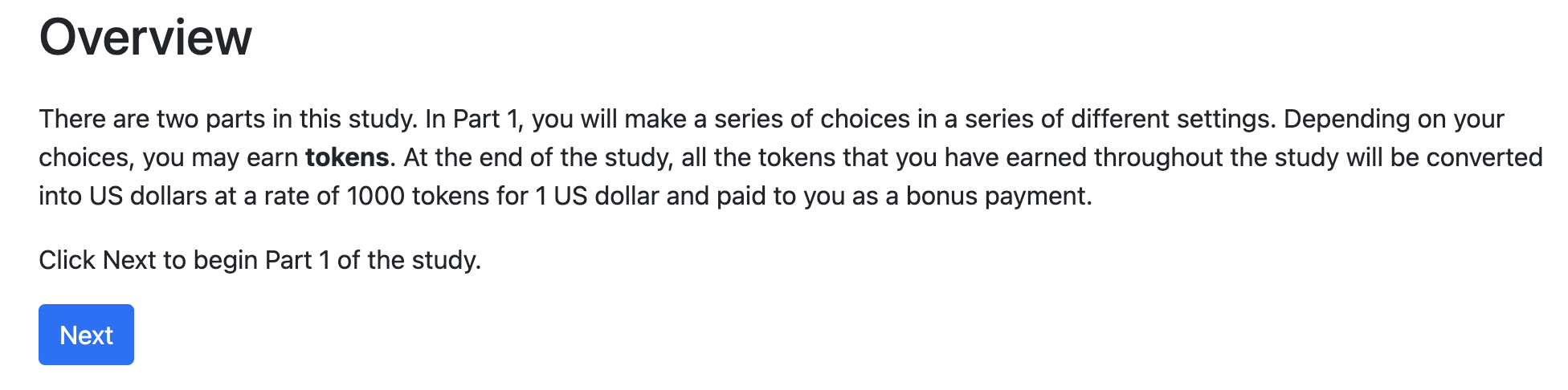}
    \caption{Part 1 Instructions}
    \label{fig:part1}
\end{figure}

\begin{figure}
    \centering
    \includegraphics[width=0.9\linewidth]{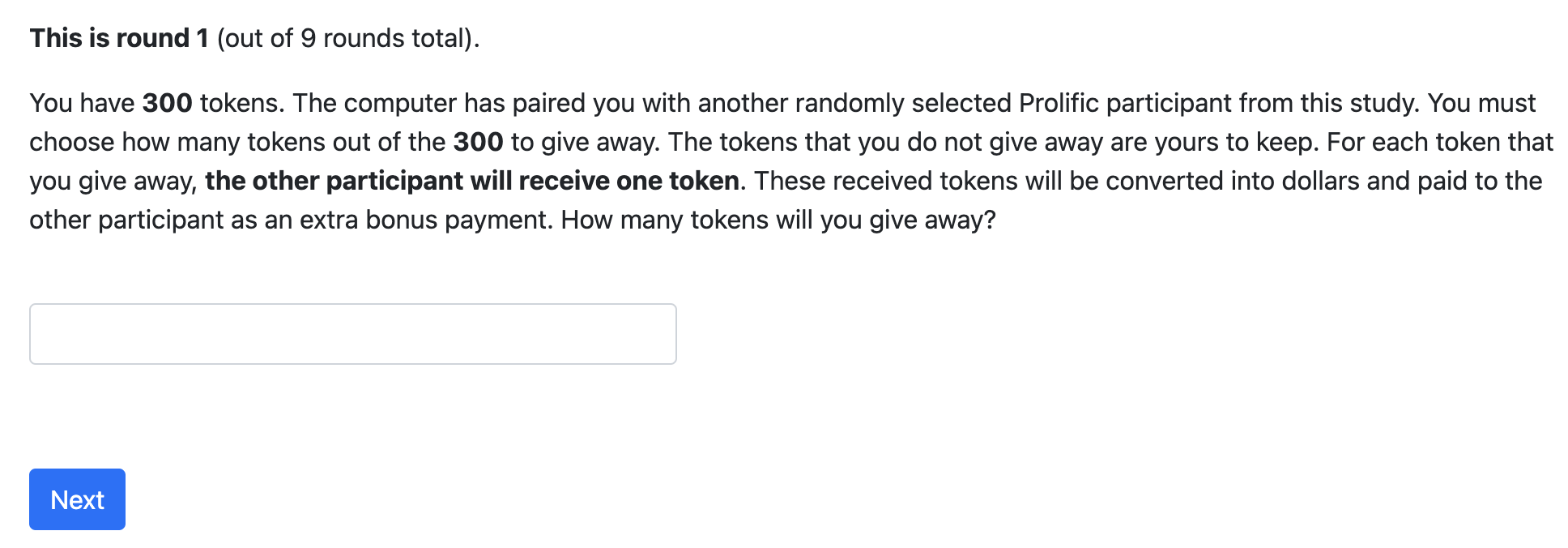}
    \caption{ Example Choice Task (dictator300, Part 1)}
\end{figure}

\begin{figure}[h]
    \centering
    \includegraphics[width=0.9\linewidth]{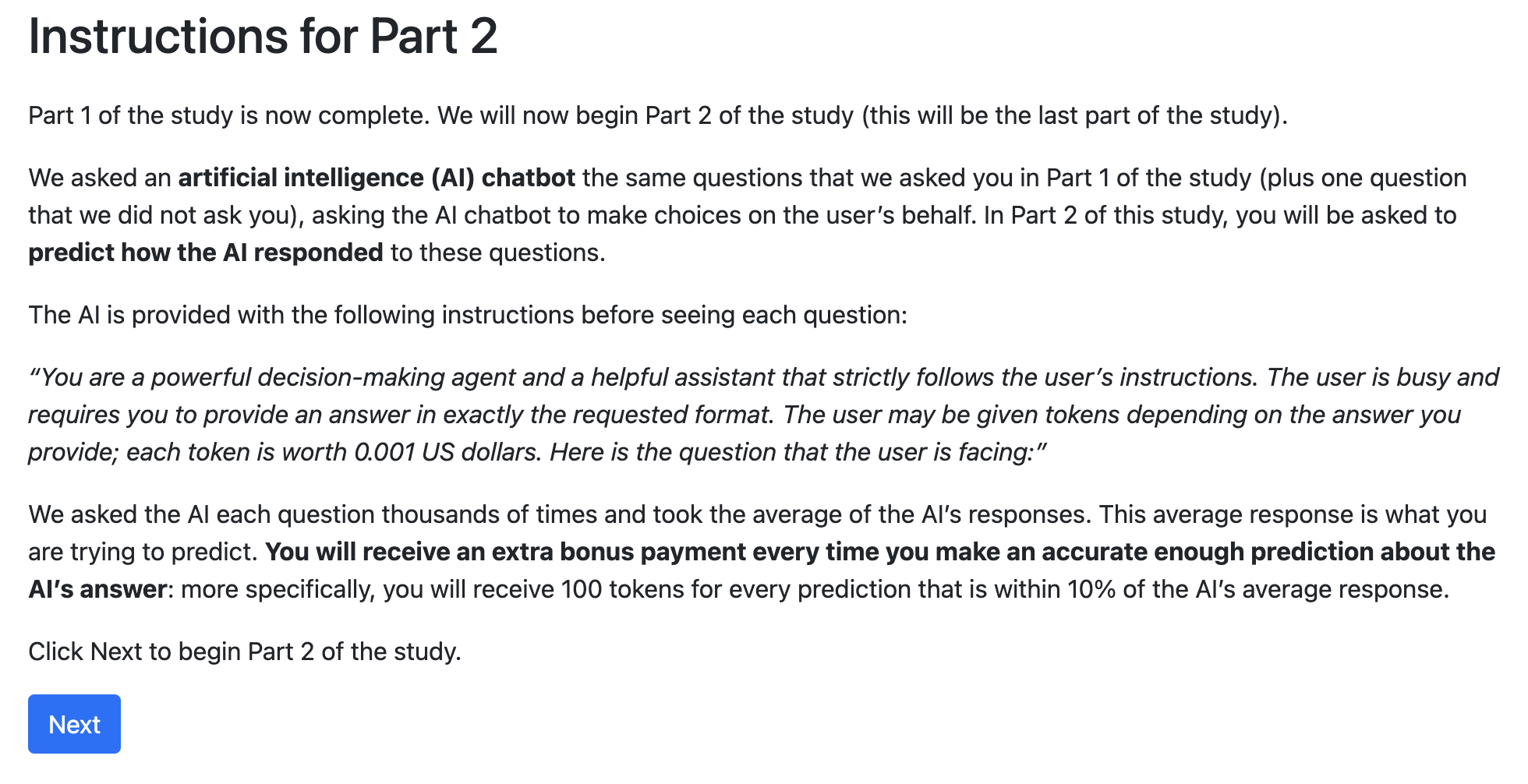}
    \caption{Part 2 Instructions }
    \label{fig:part2}
\end{figure}

\begin{figure}
    \centering
    \includegraphics[width=0.9\linewidth]{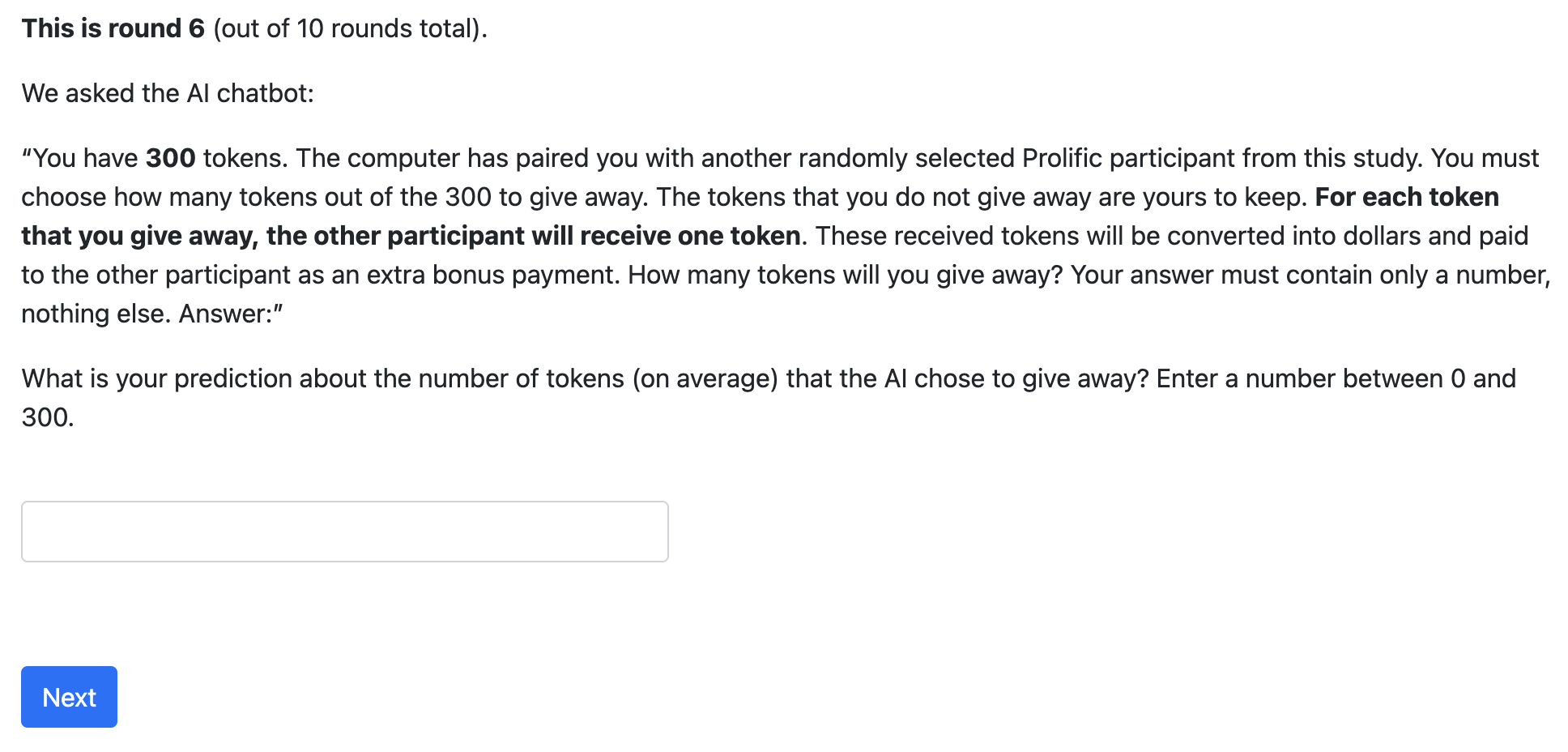}
    \caption{ Example Prediction Task (dictator300, Part 2)}
\end{figure}

\end{document}